\documentclass[twocolumn]{aastex62}
\usepackage{amsmath}
\usepackage{chngpage}

\newcommand\transa{\mbox{p-$\rm H_{2} O (2_{0,2}-1_{1,1})$}}

\newcommand\transe{\mbox{p-$\rm H_{2} O (4_{2,2}-4_{1,3})$}}

\newcommand\transg{\mbox{p-$\rm H_{2} O (2_{1,1}-2_{0,2})$}}

\newcommand\ratio{$\rm L_{H_{2}O}$/$\rm L_{FIR}$}
\newcommand{\Lsol}{L$_{\odot}$}

\newcommand\lir{$\rm L_{IR}$}
\newcommand\lfir{$\rm L_{FIR}$}
\newcommand\lwater{$\rm L_{H_{2}O}$}

\newcommand\water{\rm H$_{2}$O}
\newcommand\tdust{T$\rm _{d}$}

\def\cii {[C{\scriptsize II}]}
\def\ci {[C{\scriptsize I}]}

\def\Illinoisa{Department of Astronomy, University of Illinois at Urbana-Champaign, 1002 West Green St., Urbana, IL 61801, USA}
\def\Illinoisb{Department of Physics, University of Illinois at Urbana-Champaign, 1110 W Green St Loomis Laboratory, Urbana, IL 61801, USA}
\def\Illinoisc{National Center for Supercomputing Applications, University of Illinois at Urbana-Champaign, 1205 W. Clark St., Urbana, IL 61801, USA}
\def\Texas{Department of Astronomy, University of Texas at Austin, 2515 Speedway Stop C1400,Austin, TX 78712, USA}
\def\Arizona{Steward Observatory, University of Arizona, 933 North Cherry Avenue, Tucson, AZ 85721, USA}
\def\Merseille{Aix Marseille Univ., Centre National de la Recherche Scientifique, Laboratoire d’Astrophysique de Marseille, Marseille, France}
\def\UFlorida{Department of Astronomy, University of Florida, Gainesville, FL 32611, USA}
\def\FII{University of Florida Informatics Institute, 432 Newell Drive, CISE Bldg E251, Gainesville, FL 32611, USA}
\def\London{Department of Physics and Astronomy, University College London, Gower Street, London WC1E 6BT, UK}
\def\DAWN{Cosmic Dawn Center (DAWN), Niels Bohr Institute, University of Copenhagen, Juliane Maries vej 30, DK-2100 Copenhagen, Denmark; DTU-Space, Technical University of Denmark, DK-2800 Kgs. Lyngby}
\def\MPIfR{Max-Planck-Institut f\"{u}r Radioastronomie, Auf dem H\"{u}gel 69 D-53121 Bonn, Germany}
\def\Flatiron{Center for Computational Astrophysics, Flatiron Institute, 162 Fifth Avenue, New York, NY 10010, USA}
\def\Diego{N\'ucleo de Astronom\'ia, Facultad de Ingenier\'ia, Universidad Diego Portales, Av. Ej\'ercito 441, Santiago, Chile}
\def\ESOGarching{European Southern Observatory, Karl Schwarzschild Stra\ss e 2, 85748 Garching, Germany}
\def\Stanford{Kavli Institute for Particle Astrophysics and Cosmology, Stanford University, Stanford, CA 94305, USA}
\def\Halifax{Department of Astronomy and Physics, Saint Mary's University, Halifax, NS, B3H 3C3, Canada}
\def\Dalhousie{Department of Physics and Atmospheric Science, Dalhousie University, Halifax, NS, B3H 4R2, Canada}
\def\Leiden{Leiden Observatory, Leiden University, PO Box 9513, 2300 RA Leiden, The Netherlands}

\shortauthors{Sreevani J., et al.}

\begin{document}

\title{SPATIALLY RESOLVED WATER EMISSION FROM GRAVITATIONALLY LENSED DUSTY STAR FORMING GALAXIES AT $z$ $\sim$ 3}

\author[0000-0002-5386-7076]{Sreevani~Jarugula}\footnote{\href{mailto:jarugul2@illinois.edu}{jarugul2@illinois.edu}}
\affiliation{\Illinoisa}
\author[0000-0001-7192-3871]{Joaquin~D.~Vieira}
\affiliation{\Illinoisa}
\affiliation{\Illinoisb}
\affiliation{\Illinoisc}
\author[0000-0003-3256-5615]{Justin~S.~Spilker}
\affiliation{\Texas}
\author{Yordanka~Apostolovski}
\affiliation{\Diego}
\author[0000-0002-6290-3198]{Manuel~Aravena}
\affiliation{\Diego}
\author[0000-0002-3915-2015]{Matthieu~B\'ethermin}
\affiliation{\Merseille}
\author[0000-0002-6637-3315]{Carlos~de~Breuck}
\affiliation{\ESOGarching}
\author[0000-0002-3805-0789]{Chian-Chou~Chen}
\affiliation{\ESOGarching}
\author[0000-0001-7184-2967]{Daniel~J.M.~Cunningham}
\affiliation{\Halifax}
\affiliation{\Dalhousie}
\author[0000-0002-5823-0349]{Chenxing~Dong}
\affiliation{\UFlorida}
\author[0000-0002-2554-1837]{Thomas~Greve}
\affiliation{\London}
\affiliation{\DAWN}
\author[0000-0003-4073-3236]{Christopher~C.~Hayward}
\affiliation{\Flatiron}
\author[0000-0002-8669-5733]{Yashar~Hezaveh}
\affiliation{\Stanford}
\author[0000-0002-4208-3532]{Katrina~C.~Litke}
\affiliation{\Arizona}
\author[0000-0003-2385-6904]{Amelia~C~Mangian}
\affiliation{\Illinoisa}
\author[0000-0002-7064-4309]{Desika~Narayanan}
\affiliation{\UFlorida}
\affiliation{\DAWN}
\affiliation{\FII}
\author[0000-0001-7946-557X]{Kedar~Phadke}
\affiliation{\Illinoisa}
\author[0000-0001-7477-1586]{Cassie~A.~Reuter}
\affiliation{\Illinoisa}
\author[0000-0001-5434-5942]{Paul~Van der Werf}
\affiliation{\Leiden}
\author[0000-0003-4678-3939]{Axel~Wei{\ss}}
\affiliation{\MPIfR}


\begin{abstract}
\noindent Water (\water), one of the most ubiquitous molecules in the universe, has bright millimeter-wave emission lines easily observed at high-redshift with the current generation of instruments. 
The low excitation transition of \water, \transa\ ($\nu_{\rm rest}$ = 987.927 GHz) is known to trace the far-infrared (FIR) radiation field independent of the presence of active galactic nuclei (AGN) over many orders-of-magnitude in FIR luminosity (\lfir). This indicates that this transition arises mainly due to star formation. 
In this paper, we present spatially ($\sim$0.5$\arcsec$ corresponding to $\sim$1 kiloparsec) and spectrally resolved \mbox{($\sim$100 kms$^{-1}$)} observations of  \transa\ in a sample of four strong gravitationally lensed high-redshift galaxies with the Atacama Large Millimeter/submillimeter Array (ALMA). 
In addition to increasing the sample of luminous ($>$ $10^{12}$~\Lsol) galaxies observed with \water, this paper examines the \ratio\ relation on resolved scales for the first time at high-redshift.
We find that \lwater\ is correlated with \lfir\ on both global and resolved kiloparsec scales within the galaxy in starbursts and AGN with average \ratio$~= 2.76^{+2.15}_{-1.21}~\times10^{-5}$. We find that the scatter in the observed \ratio\ relation does not obviously correlate with the effective temperature of the dust spectral energy distribution (SED) or the molecular gas surface density. This is a first step in developing \transa\ as a resolved star formation rate (SFR) calibrator. 
\end{abstract}

\bigskip
\bigskip
\section{Introduction} \label{sec:intro}
Studies of molecules play a prominent role in understanding the physical, chemical and kinematic properties of the interstellar medium (ISM) in galaxies \citep{omont07, tielens13}. 
One such molecule is \water, the third most abundant molecule in the warm dense ISM after H$_{2}$ and CO \citep{neufeld95}.
As an asymmetric \mbox{rotor} with a large electric dipole moment, \water\ has a rich and complex spectrum giving rise to emission and \mbox{absorption} lines mainly in the submillimeter (submm) and far-infrared (FIR) regime of the electromagnetic spectrum. Observations from local galaxies \citep{weiss10,vanderwerf10,rangwala11, yang13}, high redshift ultra luminous infrared \mbox{galaxies} (ULIRGs) \citep{omont13,yang16}, and active \mbox{galactic} nuclei (AGN) \citep{vanderwerf11} have shown \water\ emission to be ubiquitous with intensities as bright as CO lines. 
Modeling has shown that, in addition to infrared pumping where \water\ is excited by FIR photons, collisions also contribute to the intensities of low-excitation transitions \citep[e.g.][]{gonzalezalfonso10,gonzalezalfonso12}. 
This is best \mbox{represented} in Figure 3 from \citet{liu17} which shows the prominent \water\ lines in different ISM \mbox{components}. The low excitation lines \mbox{become} weaker or completely disappear in the warm and hot regions \mbox{($>$ 40 K)} where infrared pumping dominates over \mbox{collisions}. The higher excitation transitions which require strong far-infrared radiation density are mainly found in the hotter regions \mbox{($100$$-$$200$ K)} of the galaxy. The \mbox{cascading} emission lines, \transa\ \mbox{(E$_{up}$ = 100.8~K,$\nu_{\rm rest}$ = 987.927~GHz)}, \transg\ (E$_{up}$ = 137~K, $\nu_{\rm rest}$ = 752.033~GHz) and p-$\rm H_{2} O$ $(2_{2,0} - 2_{1,1})$ (E$_{up}$ = 196~K, $\nu_{\rm rest}$ = 1228.789~GHz) are pumped by 101 $\mu$m photons from the base 1$_{1,1}$ level and are primarily excited in the warm regions of the galaxy. The collisional excitation of the low lying levels (1$_{1,1}$ and 2$_{0,2}$) in optically thin or high density hot regions might also contribute to the emission of the \transa\ line. Hence, \water\ transitions probe the infrared radiation field density and physical properties of the ISM such as gas density and kinetic temperature \citep[e.g.][]{weiss10,gonzalezalfonso14b,liu17}.

Because of water vapor in the Earth's atmosphere, ground-based observations of \water\ in the local universe are nearly always impossible. The \textit{Herschel Space \mbox{Observatory}}  opened the window to multiple \water\ transitions in the local universe \citep[e.g.][]{weiss10} and \citet{yang13} demonstrated that the \mbox{luminosity} of submm \water\ lines (\lwater) is linearly correlated with the total infrared luminosity (\lir, integrated over $8$$-$$1000$~$\rm \mu m$) over three orders of magnitude in multiple transitions. This suggests that the \water\ transitions, especially \transa\ which is not affected by the presence of AGN \citep{yang13}, trace the far infrared field in star forming regions.
At high redshift, \water\ has been detected using the current generation of ground-based telescopes such as the CSO, PdBI and ALMA, as the transitions are redshifted into the transparent millimeter atmospheric windows. Strong gravitational lensing, which acts as a cosmic microscope, further boosts the flux from high-redshift sources making their detections possible. Several detections of \water\ have been reported in the literature from such lensed galaxies \citep[e.g.][]{bradford09,omont11,vanderwerf11,combes12,weiss13,omont13,bothwell13b,spilker14,yang16}.

\begin{figure}[h!]
\hspace{-1.0cm}
\includegraphics[trim={0.0cm 0.0cm 0.0cm 0.0},clip,width=0.55\textwidth]{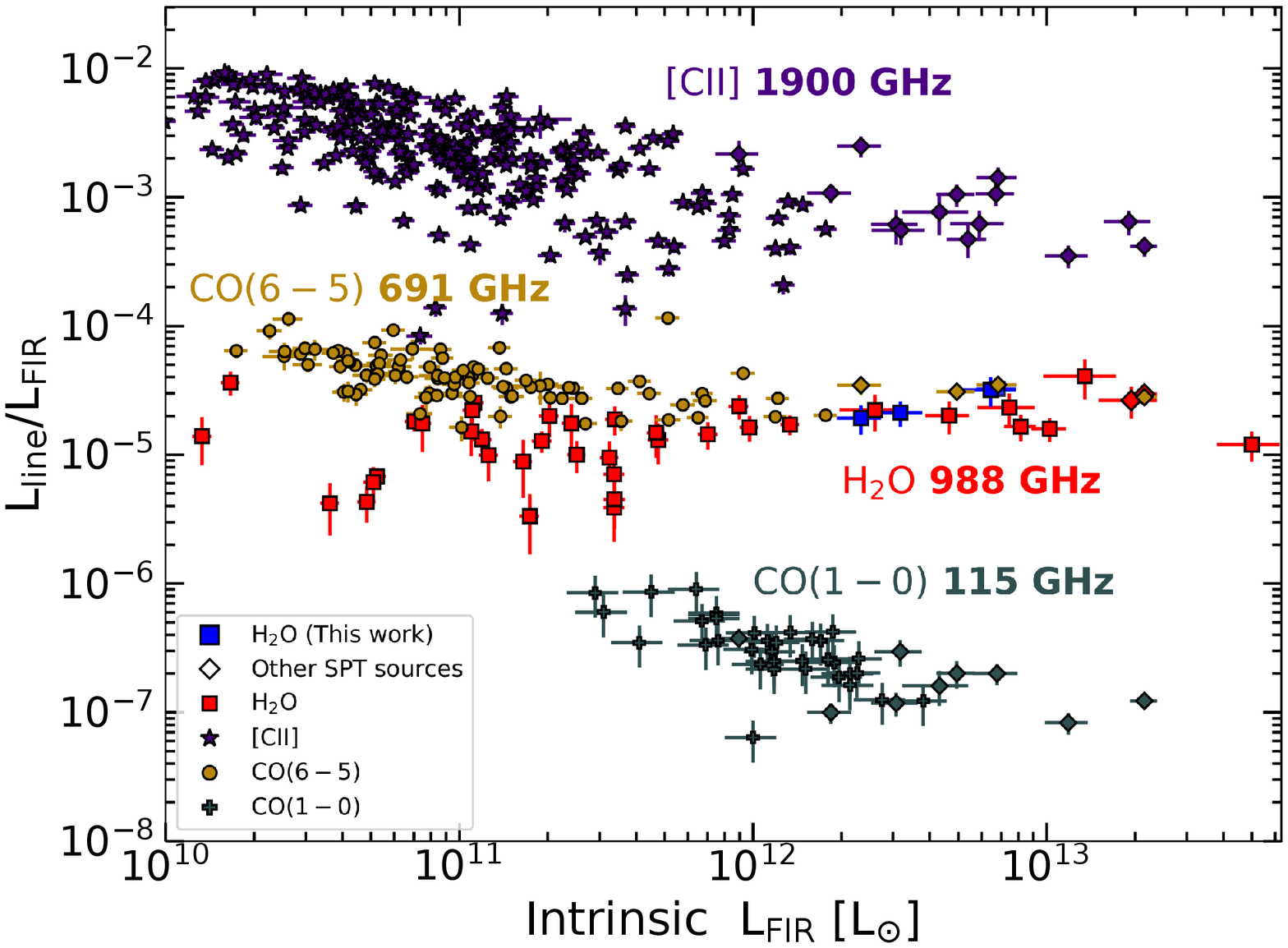}
\caption{The ratio of line to far-infrared luminosity of \transa\ (988 GHz), CO$(1-0)$ (115 GHz), CO$(6-5)$ (691 GHz) and \cii\ (1900 GHz) as a function of \lfir. The SPT sources are shown as diamonds. The \transa\ emission is described in detail in Figure \ref{mainplot}. The \water\ emission in the local galaxies is described in \citet{yang13} and the emission from high-redshift galaxies is taken from \citet{vanderwerf11}, \citet{omont13}, \citet{yang16}, \citet{apostolovski19} and this paper.
CO$(1-0)$ emission from local ULIRGs is given in \citet{solomon97} and the ATCA observations of CO$(1-0)$ in SPT sources (green diamonds) is described in detail in \citet{aravena16}. CO$(6-5)$ line emission from local luminous infrared galaxies (LIRGs) and the SPT sources (golden-yellow diamonds) is from \citet{lu17} and \citet{dong19} respectively. The \cii\ sample of LIRGs is taken from \citet{diazsantos14} and the \cii\ SPT sample represented by purple diamonds is from \citet{gullberg15}. As seen in the plot, the luminosity of CO$(1-0)$, CO$(6-5)$ and \cii\ are sublinearly correlated with \lfir\ while \transa\ is almost linearly correlated with \lfir\ especially for \mbox{\lfir\ $>$ 10$^{11.5}$ L$_{\odot}$.}}
\label{llinelfirplot}
\end{figure}

Multi-wavelength observations ranging from the UV to radio have improved our understanding of interstellar physics and the star formation rate (SFR) calibration. 
Average scaling relations from single observables are often used to estimate global SFR. Obtaining resolved SFR maps is challenging due to the difficulty in observing individual star forming regions over multiple wavelengths. Far-infrared luminosity of galaxies (\lfir, integrated over 42.5$-$122.5 $\rm \mu m$) is often used to infer SFR as it has some advantages over other indicators such as UV luminosity and recombination lines which are widely discussed in \citet{kennicutt98a} and \citet{kennicutt12}. The UV emission from young stars is a direct tracer of star formation but is highly sensitive to interstellar dust attenuation. The recombination lines such as H$\alpha$ and FIR cooling lines (e.g. \cii\ \mbox{158 $\mu$m}) originate in the ionized regions surrounding stars and are good tracers of star formation. However, these lines are affected either by dust attenuation \citep[e.g.][]{casey17} or the scatter in the estimated SFR is large \citep[e.g.][]{narayanan17,lagache18}. In contrast, \lfir\ is a good tracer of SFR at high optical depth, such as starburst galaxies where most of the UV light is re-emitted as infrared radiation. 
Although, it is widely used as a SFR calibrator in high-redshift starburst galaxies (see review by \citet{casey14}), the spectral energy distribution (SED) has to be fully sampled over the SED peak at $\lambda_{\rm rest}$$\sim$$100~\mu$m to estimate \lfir, which is observationally expensive. However, one further caveat is that infrared emission does not necessarily trace only the unobscured star formation. For instance, \lfir\ may overestimate the SFR in regions where there  are other sources of dust heating such as evolved older stars or an obscured AGN \citep[e.g.][]{kennicutt09,murphy11,hayward14}.
Longer wavelength spectral features such as \transa, which are very well correlated with \lfir\, and observable with current generation telescopes, can be used instead of \lfir\ to estimate SFR (in environments where the \lfir\ based calibration holds true). 
Figure \ref{llinelfirplot} summarizes some of the measurements from the literature and shows that \transa\ is almost linearly correlated with \lfir\ with Pearson's correlation coefficient of $\sim$0.96.  
Among the CO transitions, it has been observed that mid to high$-$J CO transitions (e.g. CO(6$-$5) and CO(7$-$6)) are a good tracer of \lir\ both in local and high-redshift (U)LIRGs \citep[e.g.][]{lu15,yang17}. However, sub-linear slopes in the \lfir-L$_{\rm CO}$ correlation arising possibly from shocks/turbulence and detached from star formation have also been discussed in high$-$J CO lines \citep[e.g.][see section \ref{subsec:co}]{greve14}. While CO is collisionally excited by H$_{2}$ molecules, \transa\ is excited by FIR photons which makes \water\ a more direct tracer of star formation. In nearby luminous galaxies, dense gas tracers such as HCN and CS are shown to be tightly correlated with \lir\ while HCO$+$ has a slightly super-linear correlation \citep[e.g.][]{gao04,zhang14}. 
\transa\ is a bright emission line (compared to HCN/HCO$+$) which is easily observable both in local and high-redshift galaxies. The linear correlation between \lwater\ and \lfir\ from Figure \ref{llinelfirplot} suggests that it is a better tracer of \lfir\ compared to other commonly observed lines such as CO(1$-$0), CO(6$-$5) and \cii. While the correlation is tight on the global integrated scales, it is unclear if this correlation breaks down on resolved scales.

In this work, we show that \lwater\ traces far infrared radiation not just at the integrated global scale \citep{yang13,omont13,yang16} but also at resolved scales within galaxies at high redshift. 
The resolution of the observations is $\sim$0.5 \arcsec\ which corresponds to $\sim$1 kiloparsec given the magnification and redshift of the sources from \citet{spilker14} (the beam resolution and the physical scale for each source are given in Table \ref{sourceprop_table} and Table \ref{tab:src_prop} respectively). 
This physical scale is only an approximation as we do not perform lens modeling in this analysis and adopt magnification values obtained from 870 $\mu$m imaging. We have selected a sample of strong-gravitationally lensed dusty star forming galaxies \mbox{(DSFGs)} discovered in the South Pole Telescope (SPT) survey \citep{vieira10,carlstrom11,mocanu13}. DSFGs host intense star formation with SFR $>$ 10 $-$ 1000 M$_\odot$/yr \citep[e.g.][]{casey14,narayanan15}. These galaxies are bright in submm wavelengths as the ultraviolet (UV) radiation from young stars is absorbed and re-radiated by the dust in FIR. Long-wavelength dust continuum observations of such galaxies have the advantage of ``negative - K correction" \citep{blain93}, where the decrease in flux due to increase in cosmological distance is compensated by the rising flux on the Rayleigh$-$Jeans side of the SED. Thus, sources of a given luminosity can be detected largely  independent of redshift. This, in addition to gravitational lensing and the power of ALMA provides enough sensitivity and resolution to investigate the correlation between \lwater\ and \lfir\ at resolved scales in star forming galaxies which we present in this paper. 

In section \ref{sec:obs}, we summarize the ALMA observations taken over two cycles and the data reduction procedure. In section \ref{sec:results}, we present the results estimating the infrared luminosity and line properties. In section \ref{sec:discussion}, we analyze the results on \lwater$-$\lfir\ correlation and the effect of physical properties on this correlation. We conclude with a summary in section \ref{sec:conclusion}. Hereafter, \water\ refers to \transa\ at 987.927~GHz and \ratio\ refers to $\rm L_{H_{2}O(2_{0,2}-1_{1,1})}/L_{FIR}$. We use Planck 2015 flat $\Lambda$CDM cosmology where $h$ = 0.677, $\Omega_{m}$ = 0.307 and $\Omega_{\Lambda}$ = 0.693 \citep{planck16cosmo}. We estimate the total infrared luminosity (\lir) as flux integrated from 8$-$1000 $\rm \mu m$ and total far-infrared luminosity (\lfir) from 42.5$-$122.5 $\rm \mu m$ in rest frame \citep{helou85}.

\section{Observations and Data Analysis} \label{sec:obs}
We choose \transa\ as it is one of the brightest \water\ transitions and has been observed to be well correlated with \lir\ \citep[e.g.][]{yang13,liu17}. This line also falls in the transparent ALMA Band 6 for the given redshift range of the sources ($z \sim$2.78 $-$ 3.37).
We observed the \transa\ 987.927 GHz line in SPT0529-54 ($z$ = 3.369), SPT0532-50 ($z$ = 3.399) and SPT0538-50 ($z$ = 2.782) with ALMA. We also include archival data on the Cloverleaf quasar, a strongly-lensed AGN at $z$ = 2.558 in this analysis. The source properties are listed in Table \ref{sourceprop_table}. 

\subsection{Sample Selection} \label{sampleselection}
The three SPT targets were selected such that they are at a similar redshift and within 10$^{\circ}$ of each other on the sky. This selection was chosen to observe the same line transition in the three galaxies and to make observations efficient for resolved ALMA Band 9 continuum observations, which where A-rated in Cycle 5, but not yet observed. All  three sources have ALMA 870 $\mu$m imaging and lens models \citep{spilker16}. \mbox{SPT0538-50} is a possible ongoing major merger as seen from dust continuum models \citep{bothwell13} and has resolved CO$(1-0)$ and CO$(3-2)$ ATCA observations \citep{aravena13,spilker15}. SPT0529-54 and SPT0532-50 have resolved CO$(6-5)$ observations  from ALMA \citep{dong19} which we make use of in this work. 
The Cloverleaf quasar (also known as H1413+117 or QSO J1415+1129) is an extensively studied strongly-lensed AGN at high redshift \citep[e.g.][]{solomon03,weiss03} which can be compared against the star forming galaxies in this sample.

\begin{table*}
\centering
\hspace{-0.8cm}
\caption{Source properties and Observations}
\label{sourceprop_table}
\begin{tabular}{c c c c c c c c c c}
\hline\hline
Source name & IAU name & RA & DEC & $z$& $\nu_{obs}^{line}$ & $\nu_{obs}^{cont}$ & Beam & $\sigma_{cont}$ & SNR$_{line}$\\
 &  &  & &   &   &  &   &  [mJy/ & \\
 &  & ( J2000 ) & ( J2000 ) &   &   [ GHz ] & [GHz] & [ $\arcsec$ ]  & beam] & \\
\hline
SPT0529-54 & SPT-S J052903-5436.6 & 05:29:03.37 & -54:36:40.30 & 3.3689 & 226.13 & 224.87& 1.10 $\times$ 1.06 & 0.17 & 7.5\\
SPT0532-50 & SPT-S J053250-5047.1 & 05:32:51.27 & -50:47:09.50 &  3.3988 &  224.59 & 223.80&0.73 $\times$ 0.56 & 0.18 & 24.4\\
SPT0538-50 & SPT-S J053816-5030.8 & 05:38:16.83 & -50:30:52.00 &  2.7817 &  260.98 & 259.32& 0.63 $\times$ 0.51 & 0.22 & 18.6\\
Cloverleaf & H 1413+117 & 14:15:46.24 & 11:29:43.68 & 2.5579 &  277.67 & 278.83&0.54 $\times$ 0.50  & 1.07 & 11.8\\

\hline\hline
\multicolumn{10}{p{\textwidth}}{NOTE. - The position (RA, DEC) and redshift ($z$) of the SPT sources are taken from ALMA 870 $\mu$m imaging in \citet{spilker16} and \citet{weiss13} respectively. The redshift of the Cloverleaf quasar is found in \citet{solomon03}. $\nu_{\rm obs}^{line}$ is the observed frequency of \transa\ transition at 987.927 GHz rest frequency. 
$\nu_{\rm obs}^{cont}$ is the frequency of the continuum.
The beam size and the sensitivity per beam in the continuum map ($\sigma_{cont}$) at $\nu_{\rm obs}$ are shown in columns 8 and 9. The peak signal-to-noise of the \water\ line (SNR$_{line}$) with 50 kms$^{-1}$ channel resolution in all sources except SPT0529-54 with 100 kms$^{-1}$ channel width is shown in the last column.}
\end{tabular}
 \end{table*}

\begin{table*}
\centering
\caption{ALMA observations}
\label{almaobstab}
\begin{tabular}{c c c c c c c c }
\hline\hline
Cycle& Proposal ID & Source & Date & Time on source & Antennas & Baseline & PWV  \\
 &  & &  & [ h ] & & [ m ]  & [mm]\\
 \hline
4&2016.1.01554.S & SPT0529-54 &  05 Dec, 2016 & 1.01 & 41 & 15.1 $-$ 650.3 &  1.57 $-$ 1.84  \\
 & & SPT0532-50 &  03 Dec, 2016 & 0.38 & 40 & 15.1 $-$ 704.1  & 0.53 $-$ 0.56 \\
 & & SPT0538-50 &  05 Dec, 2016 & 0.90 & 41 & 15.1 $-$ 650.3  & 1.57 $-$ 1.78 \\
\hline
 3&2015.1.01578.S & SPT0532-50 & 22 June, 2016 & 0.41 & 38 & 15.1 $-$ 704.1  & 1.27 $-$ 1.32 \\
 & & SPT0538-50 &  22 June, 2016 & 0.74 & 38 & 15.1 $-$ 704.1 & 1.23 $-$ 1.28\\
\hline
2  &2012.1.00175.S & Cloverleaf &30 June, 2015 & 0.16 & 42 &  &   \\

\hline\hline
\multicolumn{8}{p{\textwidth}}{\hspace{+0.2in}{NOTE. - The baseline given in the table is the minimum and maximum. 
The last column shows the range of Precipitable Water Vapor (PWV) over the course of the observations.}}
\end{tabular}
\end{table*}

\subsection{Data Reduction and Imaging} \label{dataanalysis}
For SPT0532-50 and SPT0538-50, the observations are taken with ALMA in Cycle 3 (2015.1.01578.S) and Cycle 4 (2016.1.01554.S). SPT0529-54 was observed only in Cycle 4. The Cloverleaf quasar archival data is from  ALMA 2012.1.00175.S (PI: Van Der Werf). In both Cycle 3 and Cycle 4, the SPT sources are observed in two basebands each with 2 GHz bandwidth and 240 channels with a channel resolution of 1.875 MHz. The \transa\ line and continuum in the Cloverleaf was taken over a continuous bandwidth with two basebands each 2 GHz wide, 128 channels and channel resolution of 15.625 MHz. The observations are summarized in Table \ref{almaobstab}.

\begin{figure*}[h!t]
\begin{tabular}{ccc}
\hspace*{-1.0cm}
	\includegraphics[trim={0.0cm 0 0.0cm 0},clip,width=0.272\textwidth]{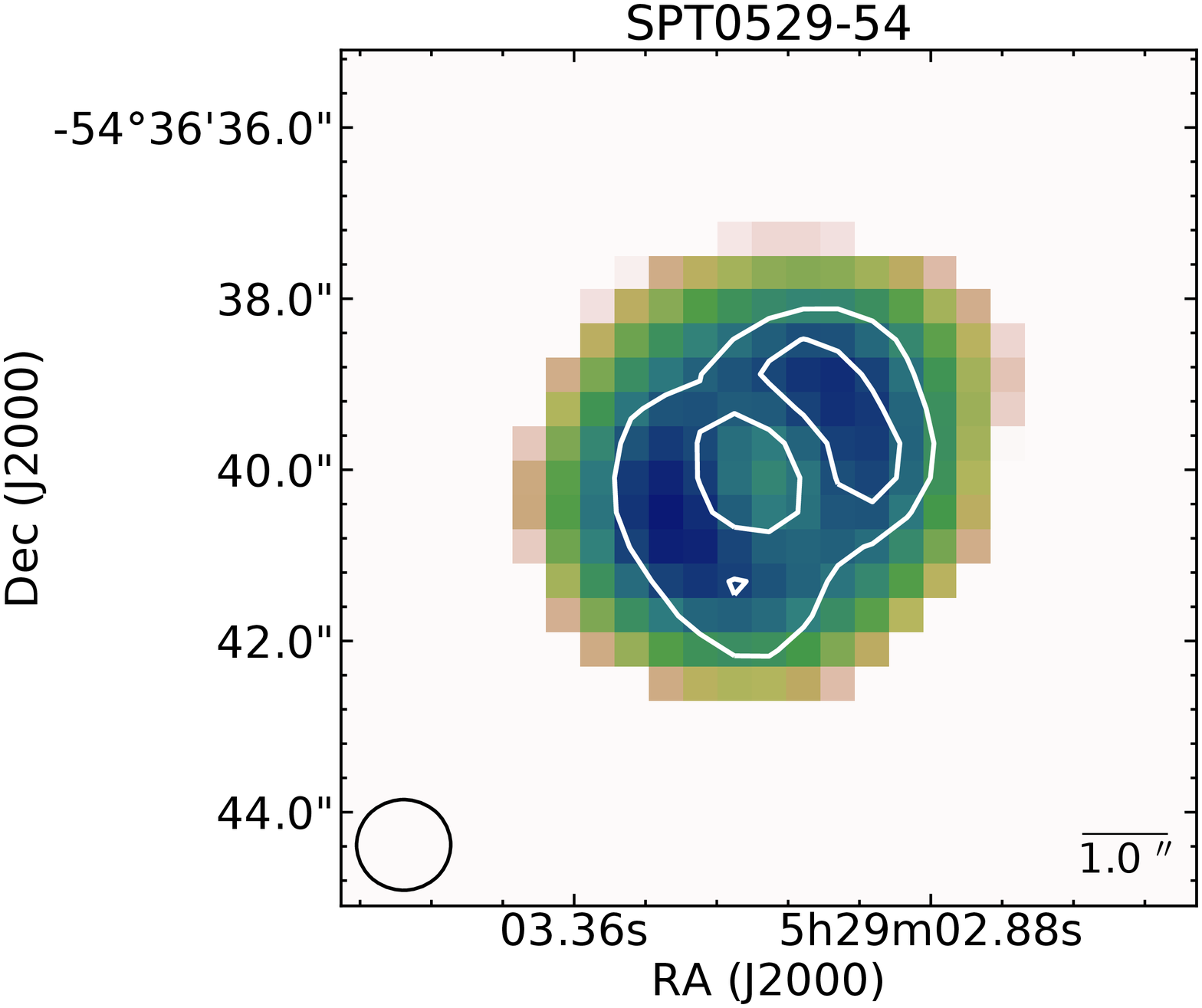}
	\hspace{0.0 cm}
	\includegraphics[trim={1.18cm 0 0.0cm 0},clip,width=0.26\textwidth]{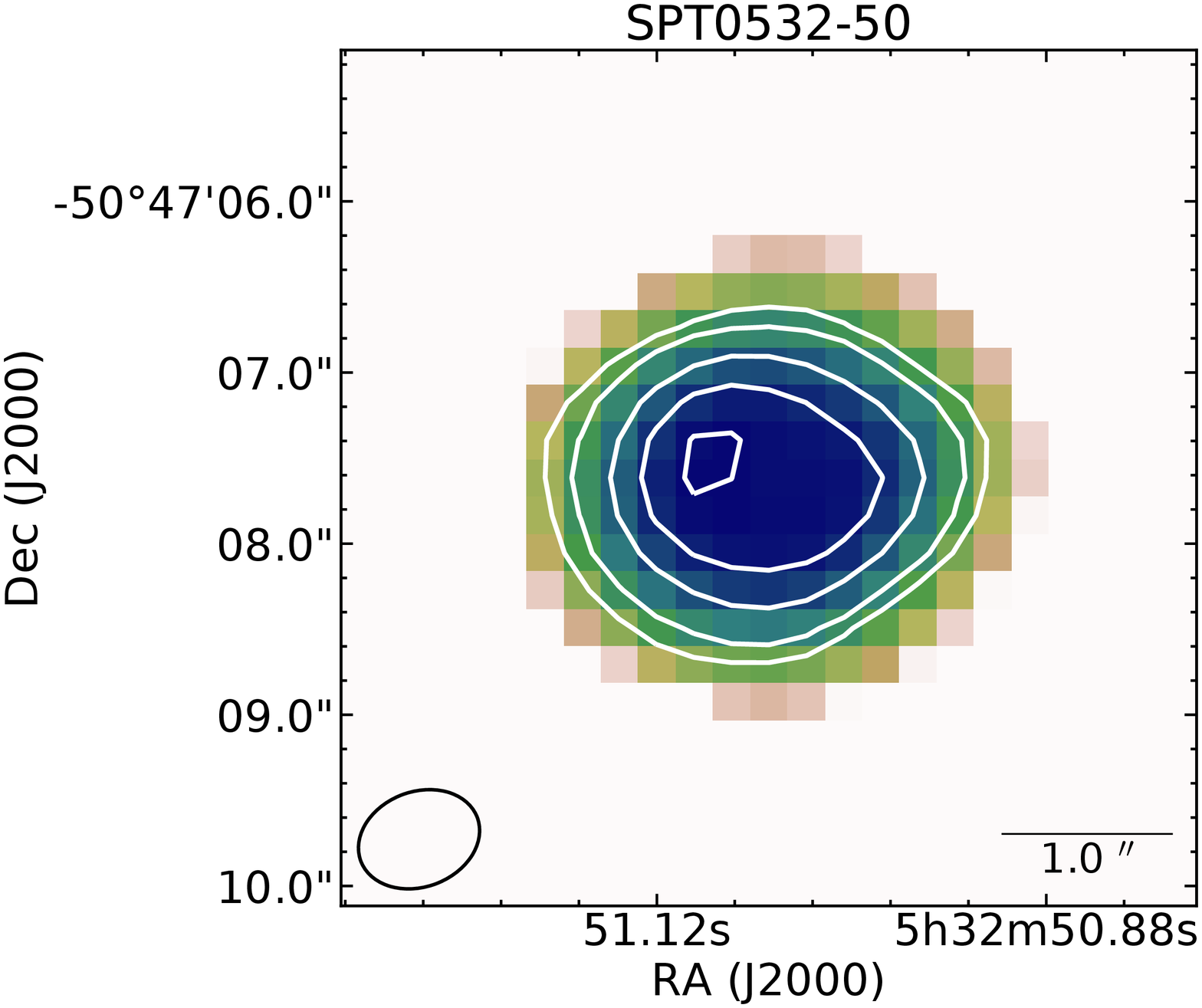}
	\hspace{0.0 cm}
	\includegraphics[trim={1.18cm 0 0.0cm 0},clip,width=0.26\textwidth]{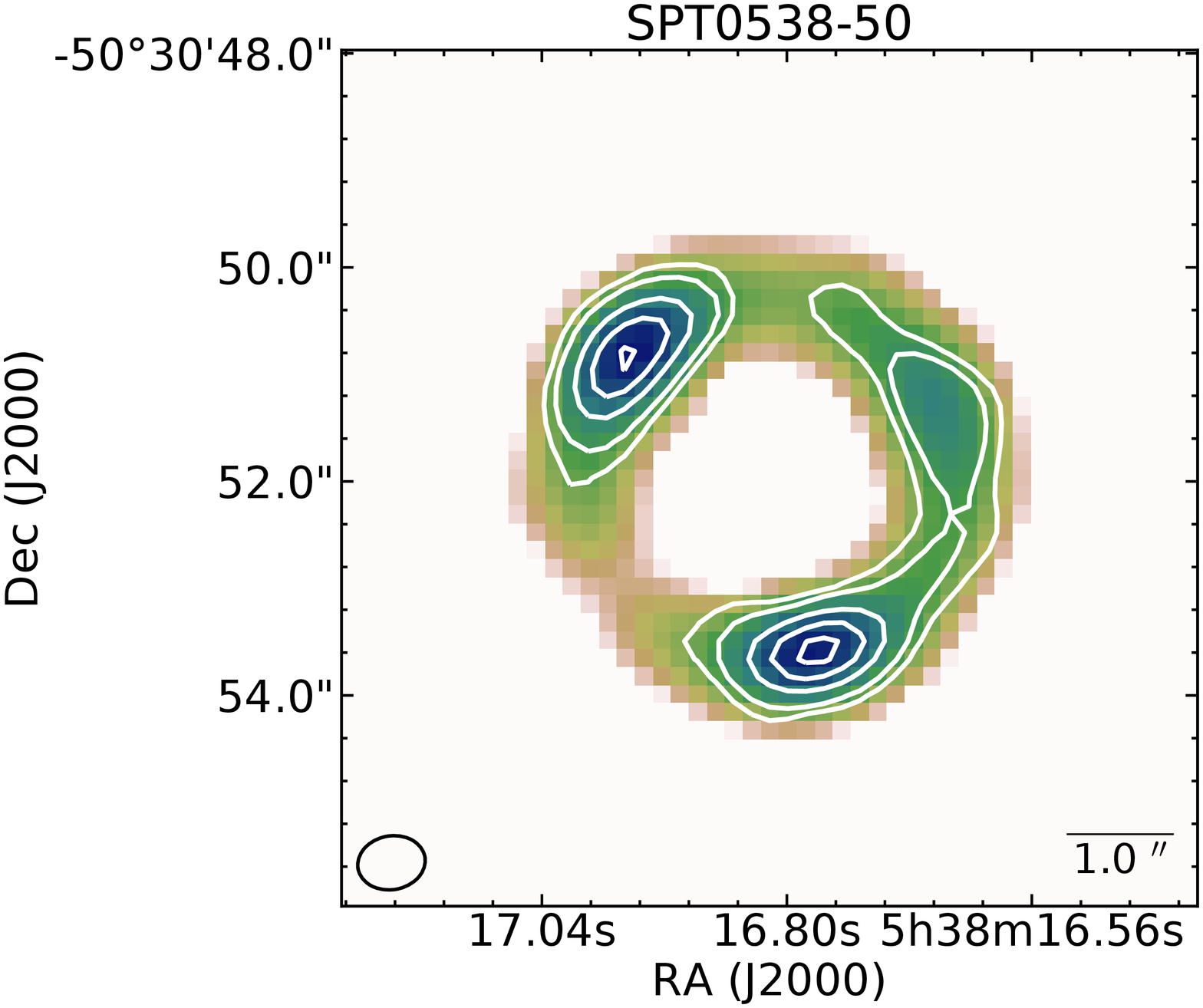}
	\hspace{0.0 cm}
	\includegraphics[trim={1.18cm 0 0.0cm 0},clip,width=0.26\textwidth]{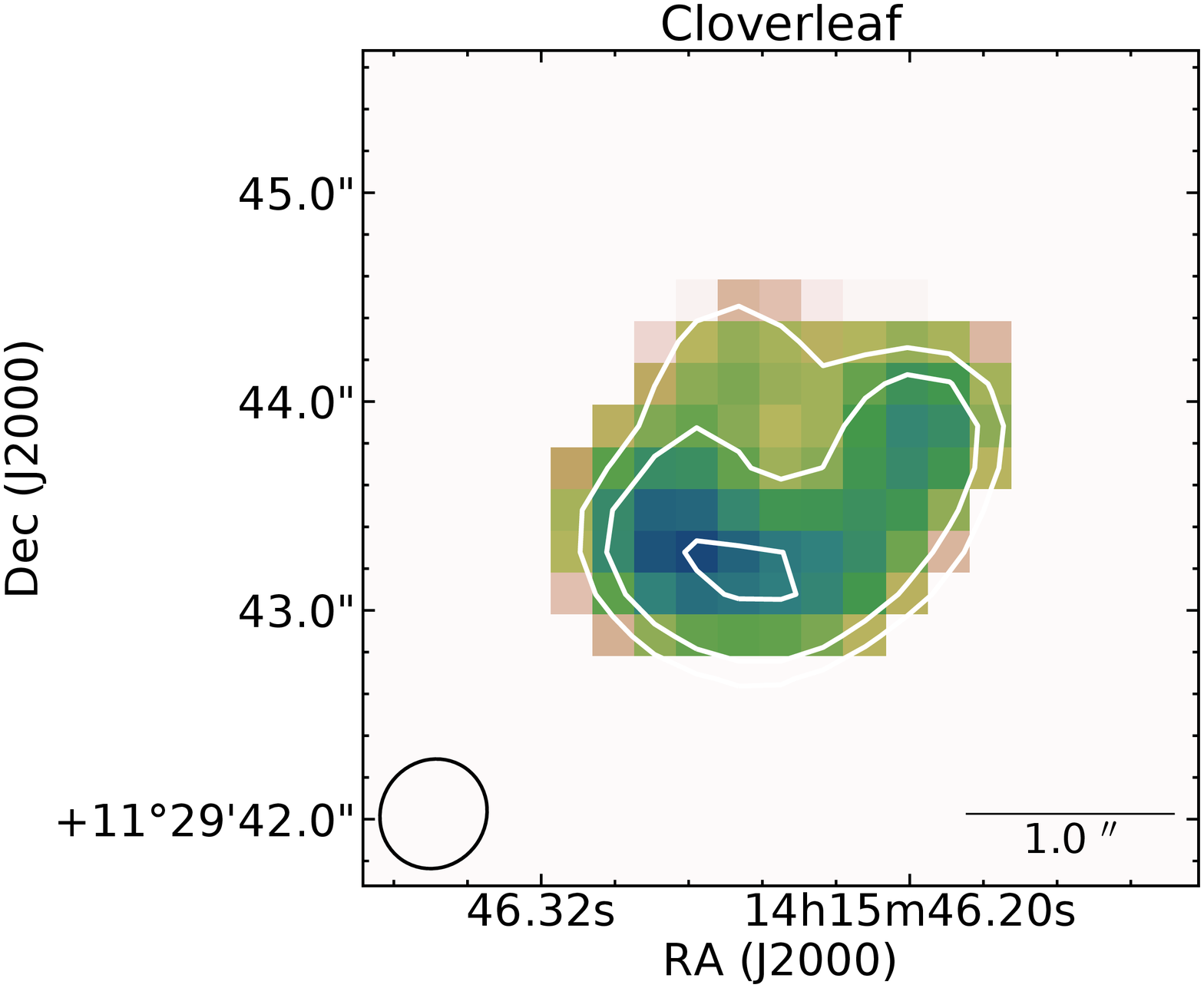}\\

\hspace*{-1.5cm}
	\hspace{1.0 cm}
	\includegraphics[trim={0.35cm 0 0.0cm 0},clip,width=0.26\textwidth]{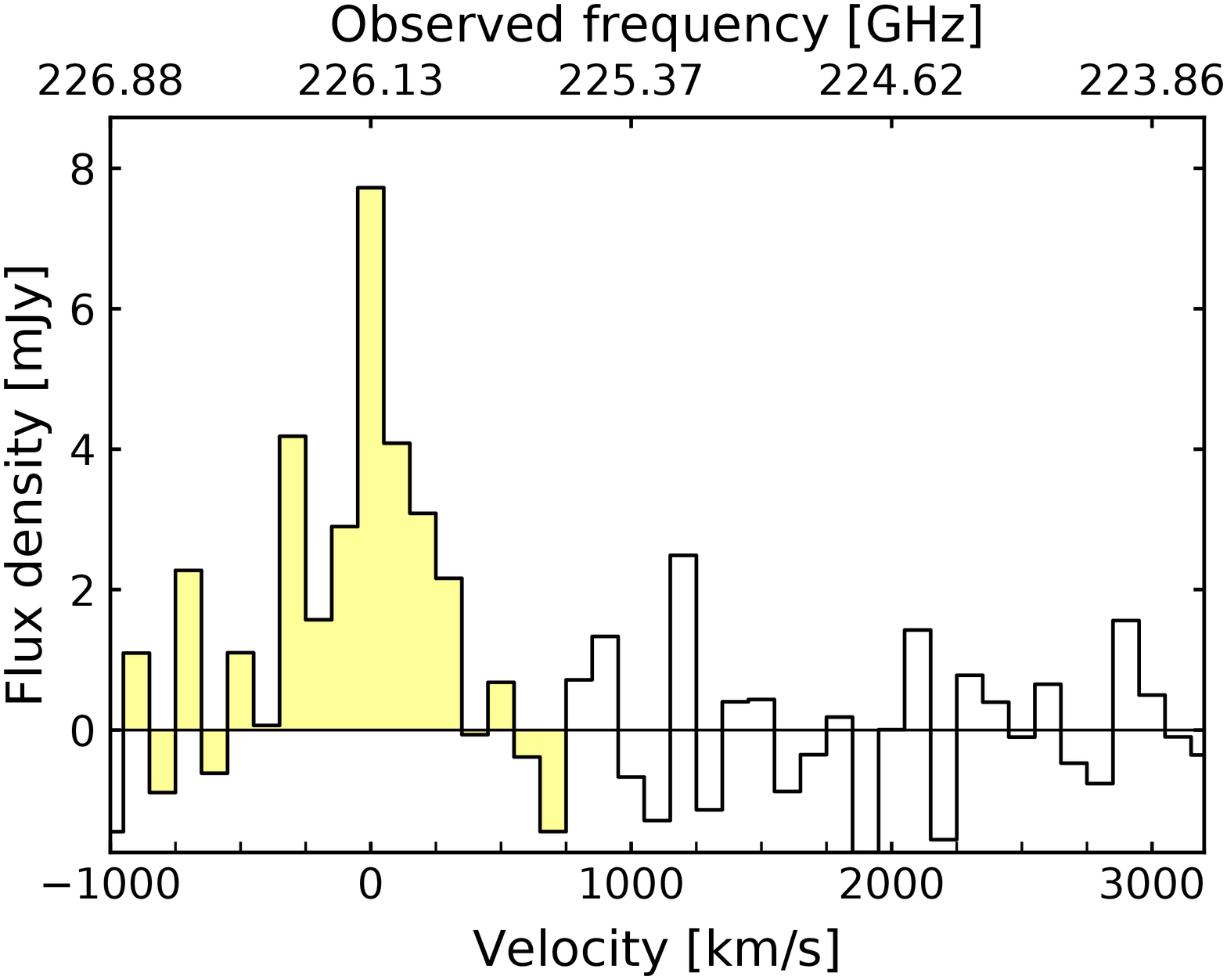}
	\hspace{0.0 cm}
	\includegraphics[trim={1.5cm 0 0.0cm 0},clip,width=0.25\textwidth]{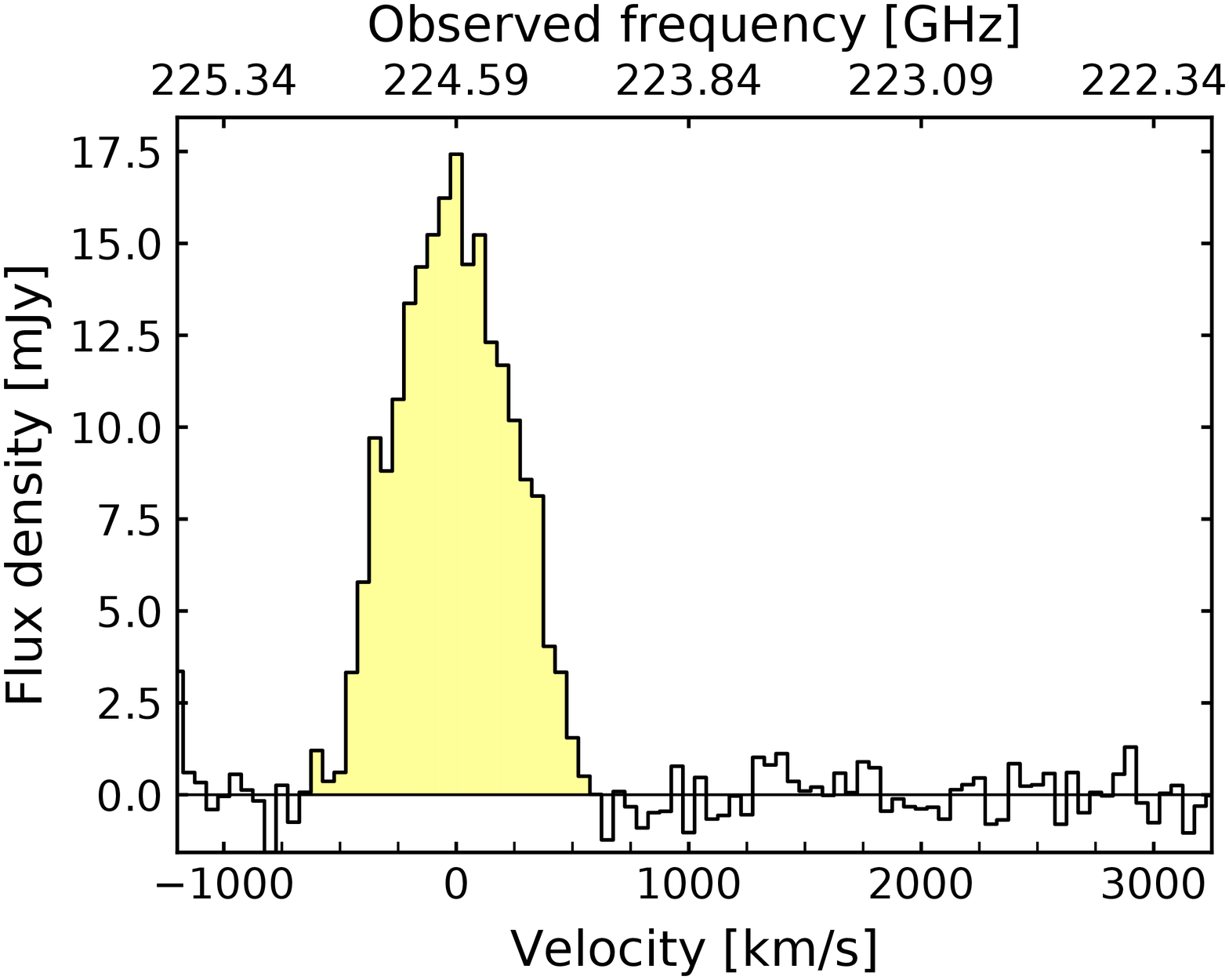}
	\hspace{0.0 cm}
	\includegraphics[trim={1.5cm 0 0.0cm 0},clip,width=0.25\textwidth]{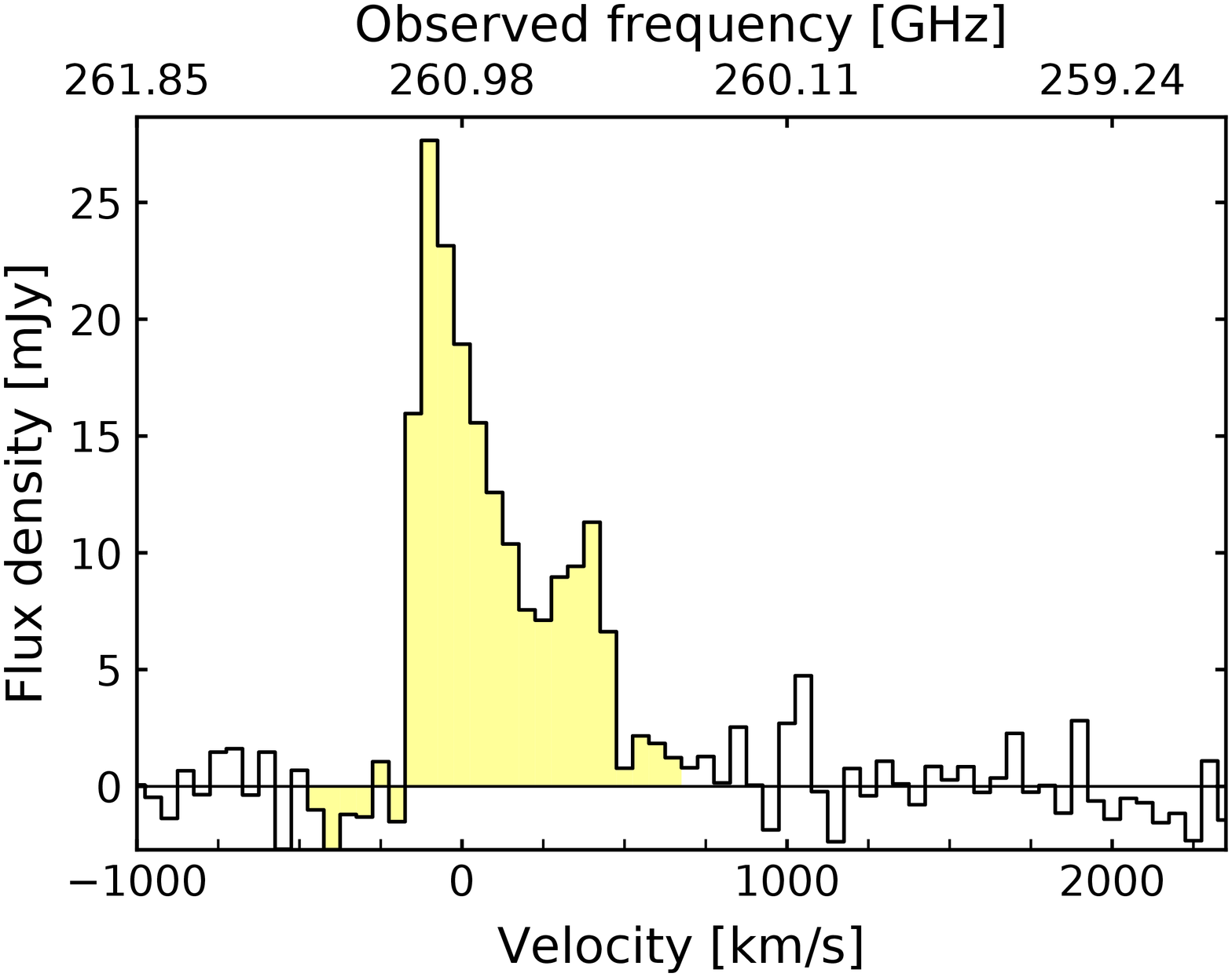}
	\hspace{0.0 cm}
	\includegraphics[trim={1.5cm 0 0.0cm 0},clip,width=0.25\textwidth]{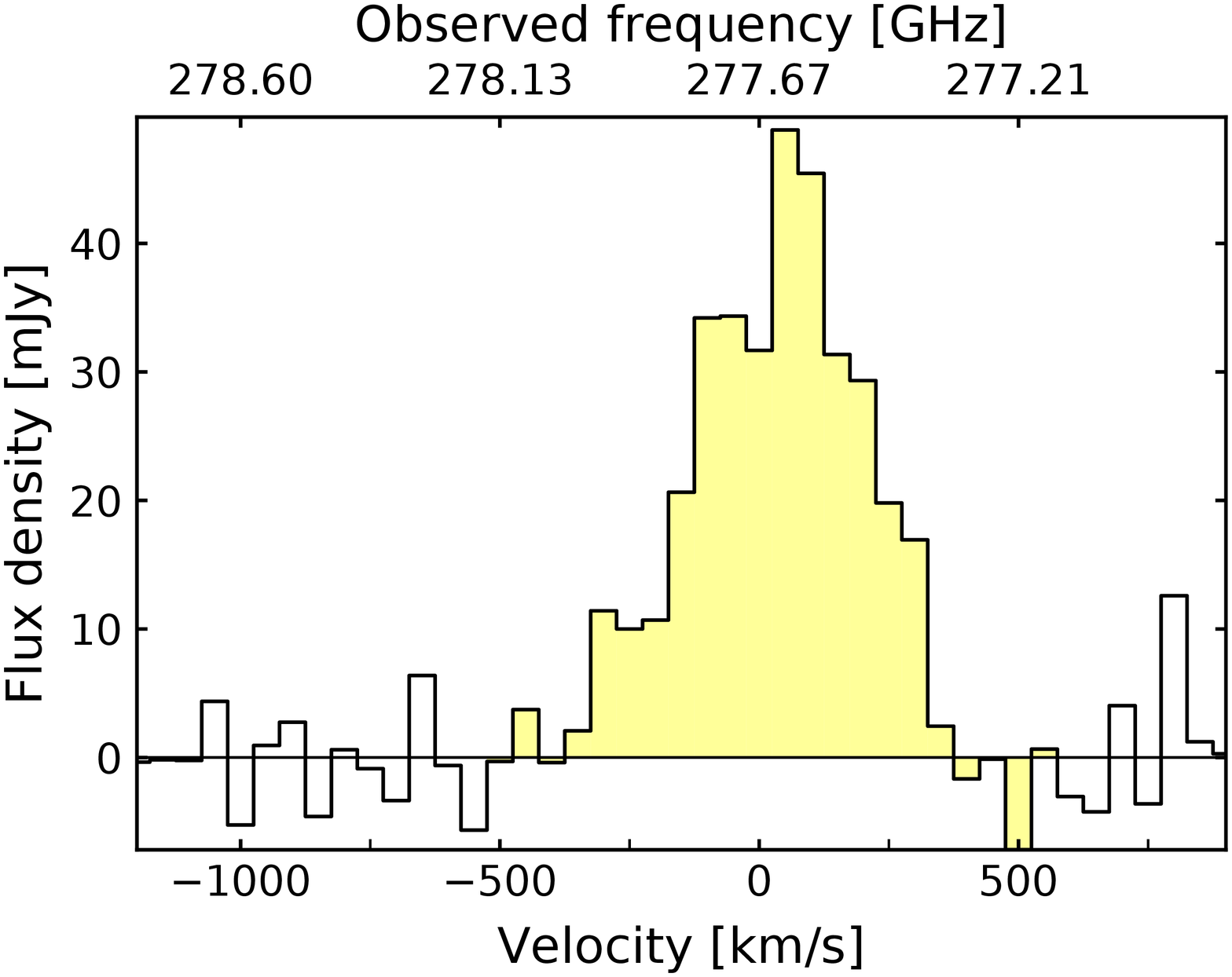}
	\end{tabular}
\caption{\textbf{Top:} The continuum is shown as background in log scale with the minimum pixel value as 3$\sigma_{cont}$ of the continuum map and the moment 0 contours of the \water\ emission are overlaid in white. The contours are at [3,5,10,15 ...] x $\sigma$ where $\sigma$ is the RMS noise in the \water\ moment 0 map. The synthesized beam of the continuum image is shown in the lower left corner and the spatial scale bar of 1.0$^{\prime\prime}$ shown in lower right. \textbf{Bottom:} The spatially integrated spectrum of \transa\ transition with 50 kms$^{-1}$ spectral resolution in all sources except SPT0529-54 with 100 kms$^{-1}$ resolution. The colored region shows (line centre - 3$\sigma_{\nu}$) kms$^{-1}$ to (line centre + 3$\sigma_{\nu}$) kms$^{-1}$ (definition given in the text.)}
\label{spectrumfig}
\end{figure*}

All the data are calibrated using the ALMA pipeline for the respective cycles. We inspected the quality of the reduction manually and found no major problems. The data were reduced and imaged using the Common \mbox{Astronomy} Software Application package \texttt{CASA} \citep{mcmullin07}. We imaged the continuum by combining data from all the spectral windows and by excluding the line emission using the task \texttt{CLEAN}. The frequency of the continuum image is given in Table \ref{sourceprop_table}. An outer taper of 1.0$^{\prime\prime}$ and 0.5$^{\prime\prime}$ is applied to SPT0529-54 and the Cloverleaf respectively such that the visibilities at shorter baselines are weighted more. This increases the signal-to-noise at the expense of resolution of the image. 
A natural weighting is applied to all the sources.
All the pixels in the image plane are correlated in the interferometric data. In order to have minimum number of correlated pixels, we choose to have few pixels in a beam ($\sim$3 to 5). The continuum images are shown in Figure \ref{spectrumfig}. We create a mask such that only the pixels with signal-to-noise $\ge$ 3 in the continuum are selected. We use this mask for all the resolved analysis in this paper.

To get the spectral cubes, we use natural weighting for all the sources and 1.0$^{\prime\prime}$  and 0.5$^{\prime\prime}$  outer taper to  SPT0529-54 and the Cloverleaf respectively, same as the continuum map.
We use 50 kms$^{-1}$ velocity averaging for SPT0532-50, SPT0538-50 and the Cloverleaf and 100 kms$^{-1}$ in SPT0529-54. For the velocity integrated intensity map (moment 0), in order to increase the signal-to-noise ratio, we re-image the data to create a single wide channel which contains most of the line flux. The width of this channel is $\sim$2 x FWHM of the line (see Table \ref{tab:src_prop} for FWHM and integrated line flux values). All the moment 0 maps are imaged similar to the continuum map and the cube. The moment 0 contours are overlaid on the continuum image in Figure \ref{spectrumfig}.

It has been shown by many studies that the CO gas sizes can be larger than that of the infrared emission \citep[e.g.][]{spilker15,tadaki17,calistrorivera18,dong19}. As \water\ and CO are observed to have similar line profiles \citep[e.g.][and Appendix A3 (\ref{sec:appendix3})
]{omont13, yang16}, it is likely that \water\ and CO are both tracing similar regions. Hence, \water\ could be also more extended than dust. In our analysis, the mask selected from the continuum includes \mbox{95$-$100$\%$} of the total \water\ flux (depending on the source) and any possible additional extended emission would be small and not affect our results. Moreover, selecting a mask based on continuum is less biased because of the high signal-to-noise in every pixel in the continuum unlike the \mbox{moment 0} map.

\section{Results} \label{sec:results}
\subsection{Estimating $L_{FIR}$} \label{sec:lfir}
To fit the SED (and estimate the total \lfir) in the SPT sources, we use the unresolved millimeter and submm photometry from ALMA  (3 mm), SPT (2.0 and 1.4 mm), LABOCA (870 $\mu$m), and \textit{Herschel} (500, 350, 250, 160 and 100 $\mu$m)  \citep{vieira13,strandet16}. For the Cloverleaf, we use the unresolved photometry from \citet{weiss03} (references therein). We fit the modified blackbody function given by Equation 1 with Markov Chain Monte Carlo (MCMC) algorithm using  the \texttt{emcee} \citep{foremanmackey13} package to sample the posterior probability function:
\begin{equation}
\rm S_{\nu} = \frac{\Omega}{(1+z)^3}(B_{\nu}(T_{d}) - B_{\nu}(T_{CMB}))(1-e^{-\tau})
\end{equation} 
where $\nu$ is the rest frequency, $\Omega$ is the source solid \mbox{angle}, B$_{\nu}(\rm T_{d})$ is the Planck function estimated at dust \mbox{temperature} $\rm T_{d}$ and $\tau$ is the optical depth. At long wavelengths, $\tau$ is given by $\tau = ( \lambda / \lambda_{o})^{\beta}$ with $\lambda_{o}$ being the wavelength at which the optical depth is unity \citep[e.g.][]{draine06} and $\beta$ is the spectral index which  determines the slope of the Rayleigh$-$Jeans tail of the blackbody. This is the same method used previously in, e.g., \citet{greve12} and \citet{spilker16}.
It should be noted that this simplistic modified blackbody function applies only with the assumption that the entire source has a single \tdust\ and that the source is \mbox{uniform} \citep[e.g.][]{hayward12}. Moreover, $\beta$, $\tau$ and \mbox{temperature} distribution are degenerate \citep{papadopoulos10}.

In this analysis we consider photometry points $\ge$~50~$\mu$m rest frame. The $\beta$ value converges to $\sim$2 for the SPT sample which has a well sampled SED. For the SPT sources and the Cloverleaf, we fix $\beta$ to 2 and let the amplitude ($\rm \Omega/(1+z)^3$), $\rm T_{d}$ and $\lambda_{o}$ vary. 
To investigate \ratio\ at resolved scales, the \lfir\ for each pixel is obtained by scaling the \lfir\ of the entire source using the flux contribution of each pixel  to the total continuum flux i.e. 
\begin{equation}
\rm L_{FIR}^{\textit{i}} = \frac{S^{\textit{i}}}{S} L_{FIR}
\end{equation} 
where $i$ denotes $i^{th}$ pixel and S is the continuum flux obtained by combining all the spectral windows from the observations excluding the line. It should be noted that the resolved \lfir\ is estimated by scaling the continuum which is not in the FIR regime i.e. not around the peak of the SED (the frequencies of the continuum are given in Table \ref{sourceprop_table}). Hence, the variations in dust temperature, optical depth, \textit{et cetera}, across the source are not taken into account in this analysis, and we assume that the galaxies have uniform temperature and opacity distribution. Improvement on this assumption would require spatially resolved continuum observations that sample the peak of the dust SED at rest-frame $\sim$100~$\mu$m (which as of this date have been approved, but not observed).

\subsection{Literature Sample} \label{literature_sample}
We draw our sample of sources detected at both low and high redshift, in 
\transa, from the literature. The local sample is drawn from the \textit{Herschel} Science Archive \citep{yang13} and the AGN in this sample are identified in \citet{koss13}. The FIR luminosity in the local galaxies is estimated by fitting 60 and 100 $\mu$m photometry \citep{sanders03} with a modified blackbody as discussed in Section \ref{sec:lfir}. We fix $\beta$ = 2.0 to be consistent with the SPT sources and $\lambda_{0}$ to 100 $\mu$m \citep{draine06} as there are only two photometry points available.

The high-redshift ULIRGs include SPT0346-52 \\ \citep{apostolovski19}, SPT0125-47 (Appendix A2 [\ref{sec:appendix2}]), HFLS3 \citep{riechers13}, APM08279+5255 \citep{vanderwerf11}, G12.v2.30, NBV1.78, SDP17b, NAV1.195 and SDP11 \citep{omont13,yang16}. \lfir\ in HFLS3 is estimated using photometry taken from \citet{riechers13} and magnification from \citet{cooray14}. In the quasar APM08279+5255, \lfir\ is taken from \citet{weiss07,beelen06} and magnification from \citet{riechers09}. \lfir\ in all the other sources (except SPT sources) are estimated similarly as for the local galaxies (fixing $\beta$ and $\lambda_{0}$) using 250, 350, 500 and 880 $\mu$m photometry and magnification from \citet{bussmann13}. All the high-redshift ULIRG values are given in Table \ref{ulirgs_table}.

\subsection{Spectral Analysis and $L_{H_{2}O}$} \label{sec_spectrum}
To obtain the spatially integrated flux density in each of the 50 or 100 kms$^{-1}$ wide channels, we apply the continuum mask to each velocity bin of the spectral cube and integrate within the region selected. 
Selecting a mask based on the continuum flux is less biased because of the high signal-to-noise in every pixel in the continuum unlike the moment 0 map. This mask includes 95$-$100$\%$ of the total water emission.
The resultant spectra are shown in Figure \ref{spectrumfig}. We use the standard deviation of flux density in line-free channels as the error in each velocity bin. To obtain the line properties, we use the non-parametric estimation of line width described in \citet{bothwell13} which is a preferred method than fitting a simple Gaussian profile, especially if the line profile is asymmetric. The intensity weighted second moment of the spectrum is given by:
\begin{equation}
\rm \sigma_{\nu} = \frac{\int(\nu - \bar{\nu})^2 S_{\nu} d\nu}{\int S_{\nu} d\nu}
\end{equation}
where $ \rm \bar{\nu}$ is the intensity weighted frequency centroid, $\rm S_{\nu}$ is the integrated flux at frequency $\rm \nu$ and the FWHM of the line is estimated as FWHM $\sim$2.35$\sigma_{\rm \nu}$. The line properties estimated with this method are listed in Table \ref{tab:src_prop} along with the velocity integrated line flux ($\rm I_{H_{2}O}$ in \mbox{Jy kms$^{-1}$}). 
We obtain the line luminosity from the relation given in \citet{solomon05}:
\begin{equation}
\rm L_{H_{2}O}  = (1.04 \times 10^{-3}) \ I_{H_{2}O} \ \nu_{rest}\ D_{L}^{2}\ (1+z)^{-1}
\end{equation}
where \lwater\ is the total line luminosity in units of $\rm L_{\odot}$, $\rm \nu_{rest}$ is the rest frequency of the line in GHz (987.927 GHz for the \transa\ transition) and $\rm D_{L}$ is the luminosity distance to the source at a redshift z in Mpc.
We estimate \lwater\ in each pixel by using the above equation with $\rm I_{H_{2}O}^{\textit{i}}$ (the integrated line flux of the $i^{th}$ pixel taken from the moment 0 image).

\begin{table*}
\centering
\caption{Observed  continuum and line properties}
\label{tab:src_prop}
\begin{tabular}{c c c c c c c c c c}
\hline\hline
Source & $\mu$ & r$^{a}$ &S$_{cont}$ & \lfir  & S$_{\rm H_{2}O}^{Peak}$  & $\rm \Delta V_{H_{2}O}$ & $\rm I_{H_{2}O}$ &\lwater &\ratio\\
 & & [kpc] &[mJy] & [10$^{12}$ L$_{\odot}$]  & [mJy] & [kms$^{-1}$] &  [Jy kms$^{-1}$] &[10$^{8}$ L$_{\odot}$] &  [10$^{-5}$] \\
\hline

SPT0529-54 & 13.2 $\pm$ 0.8 & 2.3 &32.9 $\pm$ 0.1 & 30.84 $\pm$ 4.96 & 7.7 $\pm$ 1.0 &  675 $\pm$ 235 & 2.8 $\pm$ 0.6 & 5.94 $\pm$ 1.18 & 1.92 $\pm$ 0.49 \\
SPT0532-50 & 10.0 $\pm$ 0.6 & 1.7 &42.3 $\pm$ 0.1 & 68.33 $\pm$ 9.81 & 17.6 $\pm$ 0.6 & 475 $\pm$ 44 & 10.4 $\pm$ 0.2 & 22.25 $\pm$ 0.38 & 3.26 $\pm$ 0.47\\
SPT0538-50 & 20.1 $\pm$ 1.8 & 1.1 &58.3 $\pm$ 0.1 & 63.91 $\pm$ 10.55 & 27.7 $\pm$ 1.5 & 463 $\pm$ 144 & 8.8 $\pm$ 1.3 & 13.59 $\pm$ 2.08 & 2.13 $\pm$ 0.48\\
Cloverleaf & 11  & 0.4 & 20.1 $\pm$ 0.2 & 70.91 $\pm$ 15.47 & 49.9 $\pm$ 3.7 & 434 $\pm$ 156 & 17.0 $\pm$ 2.4 & 22.63 $\pm$ 3.17 & 3.19 $\pm$ 0.83\\

\hline\hline
\multicolumn{10}{p{\textwidth}}{NOTE. - Magnification ($\mu$) of the SPT sources are taken from ALMA 870 $\mu$m lens models \citet{spilker16}. Magnification of the Cloverleaf quasar is found in \citet{venturini03}. a) r is the effective resolution in kiloparsec achieved per beam (beam size is given in Table \ref{sourceprop_table}) in the source plane.
S$_{cont}$ is the spatially integrated continuum flux ($\ge$ 3$\sigma_{cont}$) and \lfir\ is the observed far-infrared luminosity. S$_{\rm H_{2}O}^{Peak}$ is the peak line flux and $\rm \Delta V_{\rm H_{2}O}$ is the FWHM of the spatially integrated spectrum which are derived using the non-parametric method of estimating line properties (Section \ref{sec_spectrum}). $\rm I_{H_{2}O}$ is velocity integrated line flux under the colored region shown in Figure \ref{spectrumfig}. \lwater\ is the observed \water\ luminosity. The error on \ratio\ takes into account both the error on luminosities and magnification.}
\end{tabular}
\end{table*}
 
\section{Analysis and Discussion} \label{sec:discussion}
\subsection{ $L_{H_{2}O} - L_{FIR}$ Correlation and SFR Calibration} \label{subsec:lh2olfir}
\begin{figure*}
\hspace{-0.8cm}
\includegraphics[trim={0.0cm 0.0cm 0.0cm 0.0},clip,width=1.05\textwidth]{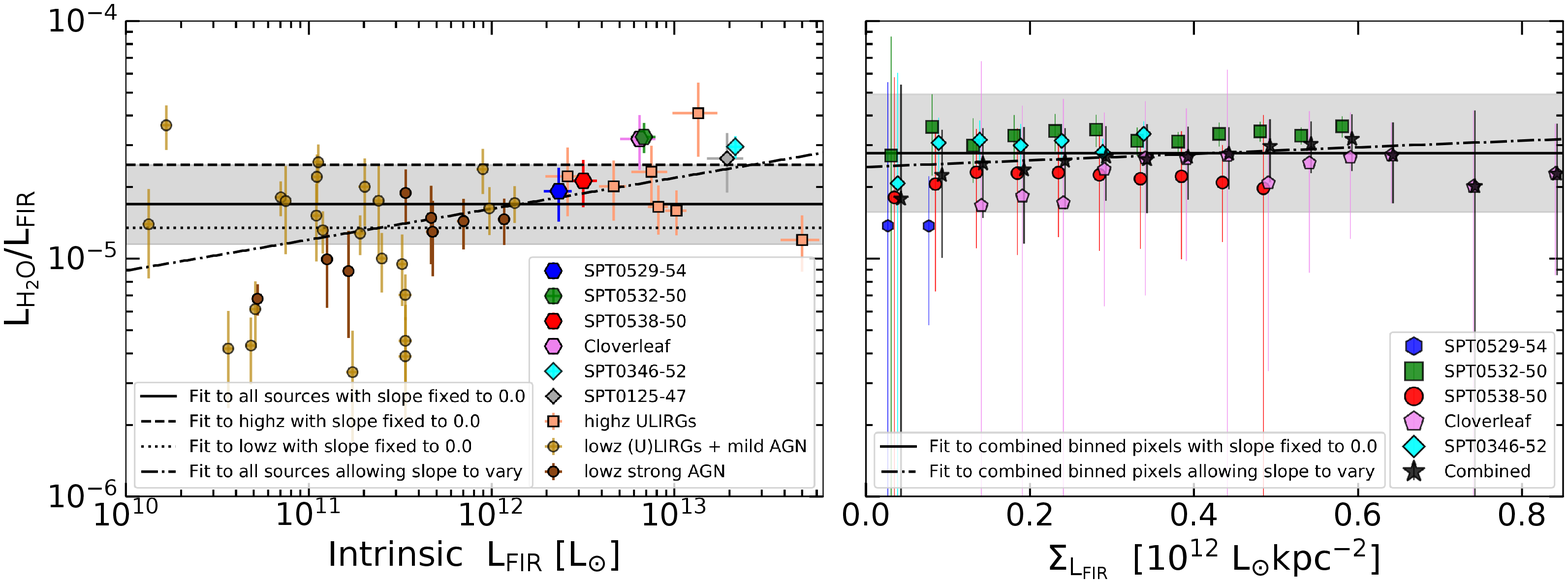}
\caption{\textbf{Left (A):} Global \ratio\ plotted as a function of spatially integrated intrinsic \lfir\ (corrected for magnification in high-redshift lensed galaxies). \lwater\ of the low-redshift LIRGs and ULIRGs (with mild and strong AGN) are from \cite{yang13}. \lwater\  of the high-redshift ULIRGs and AGN are taken from \citet{omont13}, \citet{yang16} and \citet{vanderwerf11}. The three SPT sources, SPT0529-54, SPT0532-50 and SPT0538-54 and the Cloverleaf are presented in this paper. SPT0346-52 is taken from \citet{apostolovski19} and SPT0125-47 is presented in Appendix A2 [\ref{sec:appendix2}]. The \textit{dot-dashed} line is the best fit to all the sources by allowing the slope as a free parameter.
The best fit by fixing slope to zero is shown as \textit{thick black} line and the grey region corresponds to the error on the fit. The \textit{dashed line} is a fit to the high-redshift sources and the \textit{dotted} line fits the low-redshift galaxies with a fixed slope of zero. \textbf{Right (B):} Resolved \ratio\ plotted as a function of surface brightness in units of $\rm L_{\odot} kpc^{-2}$. Each data point is the value of pixels binned within 0.05 x $\rm 10^{12}$ $\rm L_{\odot}kpc^{-2}$. The black data points are obtained by combining all the pixels from the five sources and the fit to these points by fixing the slope to zero is shown as \textit{thick black} line with the 1$\sigma$ uncertainty on the fit shown in grey. The \textit{dotted} line is the best fit by allowing the slope to vary and it is within the grey error region. As shown in the plots, \lwater\ is strongly correlated with \lfir\ both at global and resolved scales within the galaxy.}
\label{mainplot}
\end{figure*}

Using the estimated intrinsic luminosities (corrected for magnification) of FIR and \water, we plot \ratio\ as a function of \lfir\ in Figure \ref{mainplot} in both globally integrated scales and spatially resolved scales \citep[values for high-redshift galaxies are given in Table \ref{tab:src_prop} and Table \ref{ulirgs_table} and local galaxies are discussed in][]{yang13}. 
In our analysis, we assume that \water\ and FIR are co-spatially lensed and hence, the issue of differential lensing where the lensing magnification varies across the source \citep{blain99d,hezaveh12a} is not significant.

From Figure \ref{mainplot}A, the correlation between global \lwater\ and \lfir\ is slightly super-linear and shown as dot-dashed black line. A fit (including all sources from literature in the MCMC) to log$_{10}$(\lwater) and log$_{10}$(\lfir) gives:  
\begin{equation}
\rm L_{H_{2}O} \propto \rm L_{FIR}^{\hspace{0.5cm}(1.17 \pm 0.21)} 
\end{equation}
This result is similar to the conclusion presented in the literature \citep{omont13,yang16} where \ratio\ is slightly higher in luminous high-redshift galaxies when compared to less luminous local galaxies. The fit to \ratio\ with the slope fixed to zero is shown as a thick black line (the error region is shown in grey) gives \ratio\ = 1.69$^{+0.79}_{-0.54}$ $\times$ 10$^{-5}$. The fit (slope fixed to zero) to high-redshift sources is shown as dashed line and to the low-redshift galaxies is shown as dotted line. Both the lines (with their respective errors) are within the error bar of the thick black line which shows that the increase in \ratio\ in this sample might not be significant. The almost linear relation between \lwater\ and \lfir\  over more than 3 orders-of-magnitude supports the previously found conclusions in the literature that \transa\ traces \lir\ both in local galaxies and in high-redshift ULIRGs irrespective of the presence of AGN. 

It should be noted that the low-redshift sample from \citet{yang13} is not complete and the super-linear correlation between \lwater\ and \lfir, seen in Figure~\ref{mainplot}A might be real. The deficiency of \water\ in less luminous galaxies (\lfir\ $<$ 10$^{11.5}$ L$_{\odot}$) has also been observed in the compilation of fluxes presented by \citet{liu17} where a few LIRGs have \lwater\ similar to ULIRGs while the others either have lower or no \water\ detection. This effect could be arising because of \transa\ excitation requirements. \water\ molecules have to be well shielded from UV radiation to avoid dissociation but also have to reside in warm gas (not UV heated) to escape into gas phase from the grain mantles. Moreover, \transa\ requires dense gas to populate the 1$_{1,1}$ level through collisions, which is the base for 101 $\mu$m excitations. Hence, \water\ couples with far-infrared radiation strongly in such warm, dense, well shielded gas which is prevalent in galaxies with \mbox{\lfir\ $>$ 10$^{11.5}$ L$_{\odot}$,} but less so in low luminous galaxies. 
\water\ emission is also enhanced as a result of shocks or intense radiation fields prominent in starbursts. Shocks could increase the abundance of \water\ and strong radiation could lead to increase in excitations \citep[e.g][]{gonzalez10,omont11}. This might result in the slightly super-linear correlation.

Figure \ref{mainplot}B shows the correlation between \lfir\ and \lwater\ on resolved scales. In each spatial pixel, the \mbox{surface brightness} ($\rm \Sigma_{\rm L_{FIR}}$) is estimated by dividing the observed \lfir\ in each pixel by the area of the pixel in the image plane. Because lensing conserves surface brightness, the pixel-by-pixel values of $\rm \Sigma_{\rm L_{FIR}}$ do not require a lensing correction. 
\ratio\ per pixel also does not require a lensing correction as we assume that the FIR and \water\ emission are co-spatially lensed.
The data points corresponding to each source are the average value of \ratio\ in pixels binned within (0.05 x $\rm 10^{12}$) $\rm L_{\odot}kpc^{-2}$ and the error on each data point corresponds to combined error of the standard deviation of pixel values within that bin and the propagated error due to averaging. This averaging reduces the number of degenerate pixels arising from multiple images of the same region in the source due to gravitational lensing. However, pixel averaging contributes to the error on $\Sigma_{\rm L_{FIR}}$ and \ratio\ because of differential lensing within pixels in each bin. These errors are not shown in Figure \ref{mainplot}B as it is difficult to quantify without a good lens model but this does not affect the results discussed below.
The black points are obtained by combining all the pixels from all the sources. The best fit to the black points by fixing the slope to zero is given by:
\begin{equation} \label{eqn:lh2o_lfir}
\frac{\rm L_{H_{2}O}}{\rm L_{FIR}} = 2.76^{+2.15}_{-1.21} \times 10^{-5} 
\end{equation}
This fit is shown as a thick black line and the grey shaded region in Figure \ref{mainplot}B which is consistent with the global \ratio\ (Figure \ref{mainplot}A). The dotted line in Figure \ref{mainplot}B is the best fit to the resolved data by allowing the slope to vary. We obtain \ratio\ $\propto$ \lfir$^{0.14 \pm 1.44}$ which is consistent within the grey shaded region. On comparing the two fits using a F-test, we conclude that the fit with varying slope does not provide significantly more information than the fit with slope fixed to zero. We obtain a F-distribution value of $\sim$0.5 with the null hypothesis that the fit with varying slope  provides more information than the fit with slope fixed to zero and we reject this hypothesis if the F-distribution value is $>$ 0.05.

We observe that \ratio\ remains linear in SPT0532-50, SPT0538-50 and the Cloverleaf with $\Sigma_{\rm L_{FIR}}$. The correlation is not obvious in SPT0529-54 as the signal-to-noise of the \water\ emission in this source is lower than other sources (peak SNR $\sim$7.5 in 100 kms$^{-1}$ channels). This strong correlation between \lwater\ and $\rm \Sigma_{\rm L_{FIR}}$ suggests that \water\ is tracing star formation rate not just at global scales (as discussed in previous paragraph) but also at resolved scales ($\sim$1 kiloparsec) within the galaxy. 
Thus, this result demonstrates that we can use resolved \water\ as resolved SFR indicator in high-redshift intense star forming regions. However, high resolution continuum observations in the FIR regime are required to quantify temperature (and hence \lir) variations within the galaxy.

\begin{figure*}[h!t]
\hspace*{-1.0cm}
	\begin{tabular}{cccc}
	\includegraphics[trim={2.0cm 3.5 2.0cm 0},clip,width=0.24\textwidth]{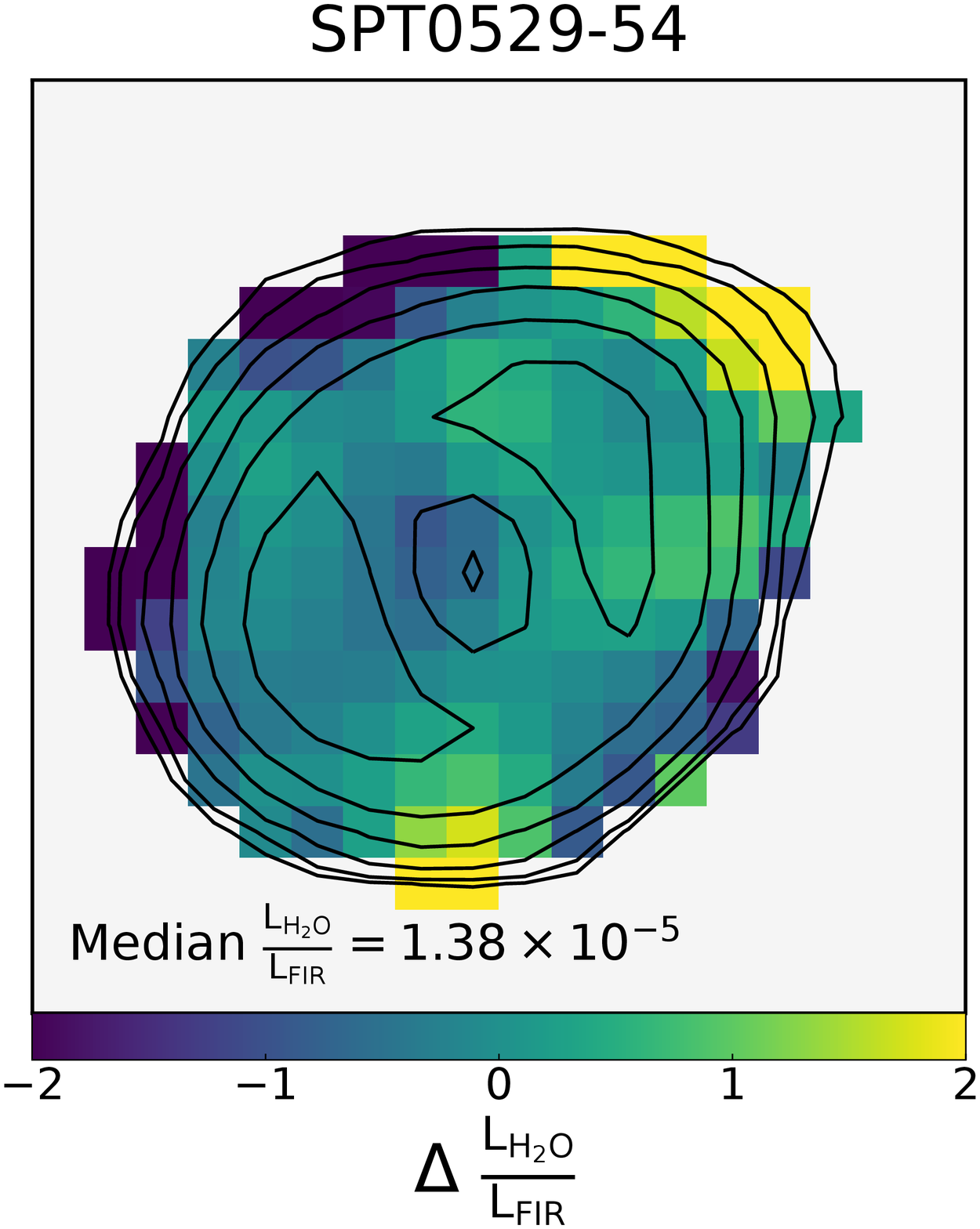}
	\hspace{0.0 cm}
	\includegraphics[trim={2.0cm 3.5 2.0cm 0},clip,width=0.24\textwidth]{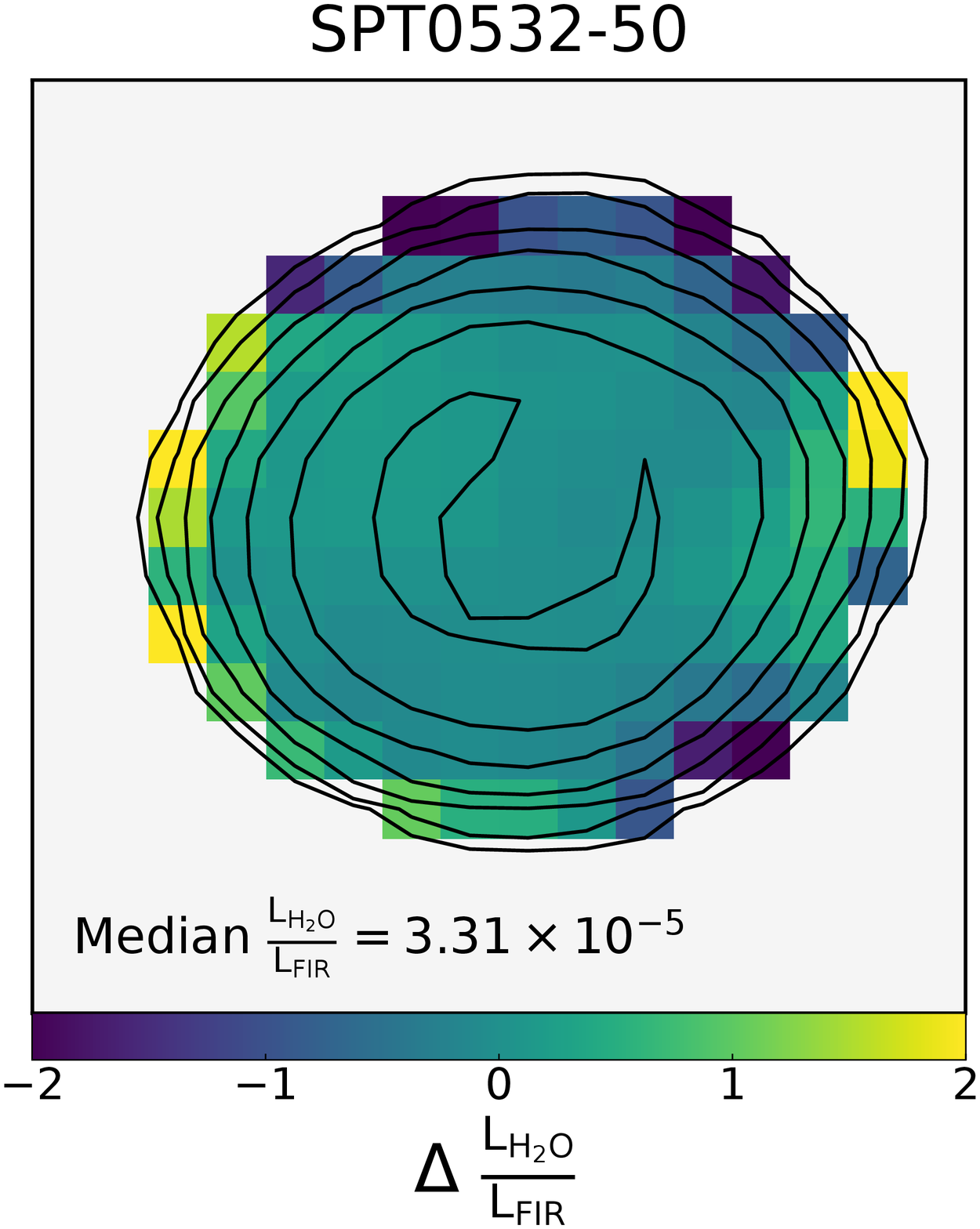}
	\hspace{0.0 cm}
	\includegraphics[trim={2.0cm 3.5 2.0cm 0},clip,width=0.24\textwidth]{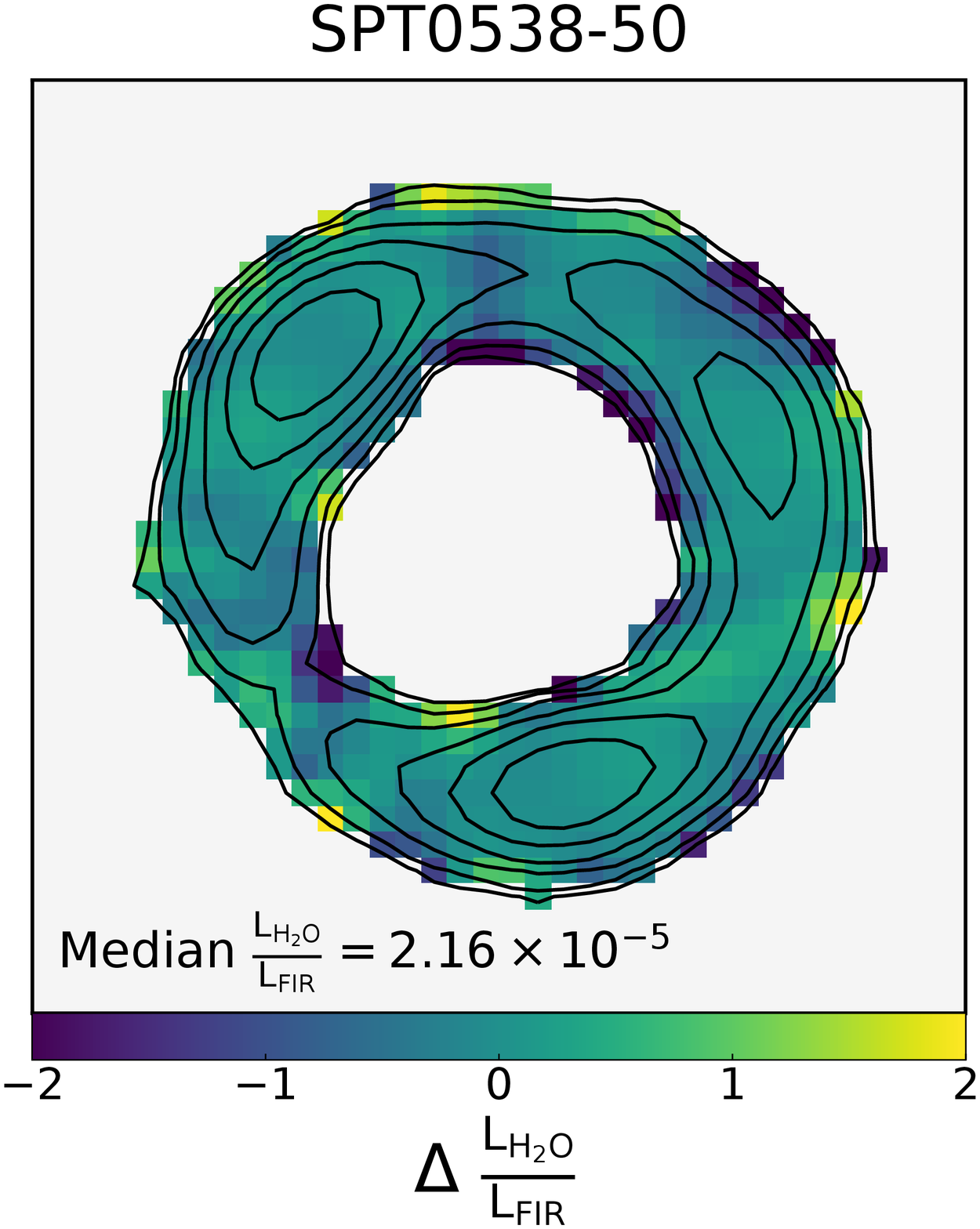}
	\hspace{0.0 cm}
	\includegraphics[trim={2.0cm 3.5 2.0cm 0},clip,width=0.24\textwidth]{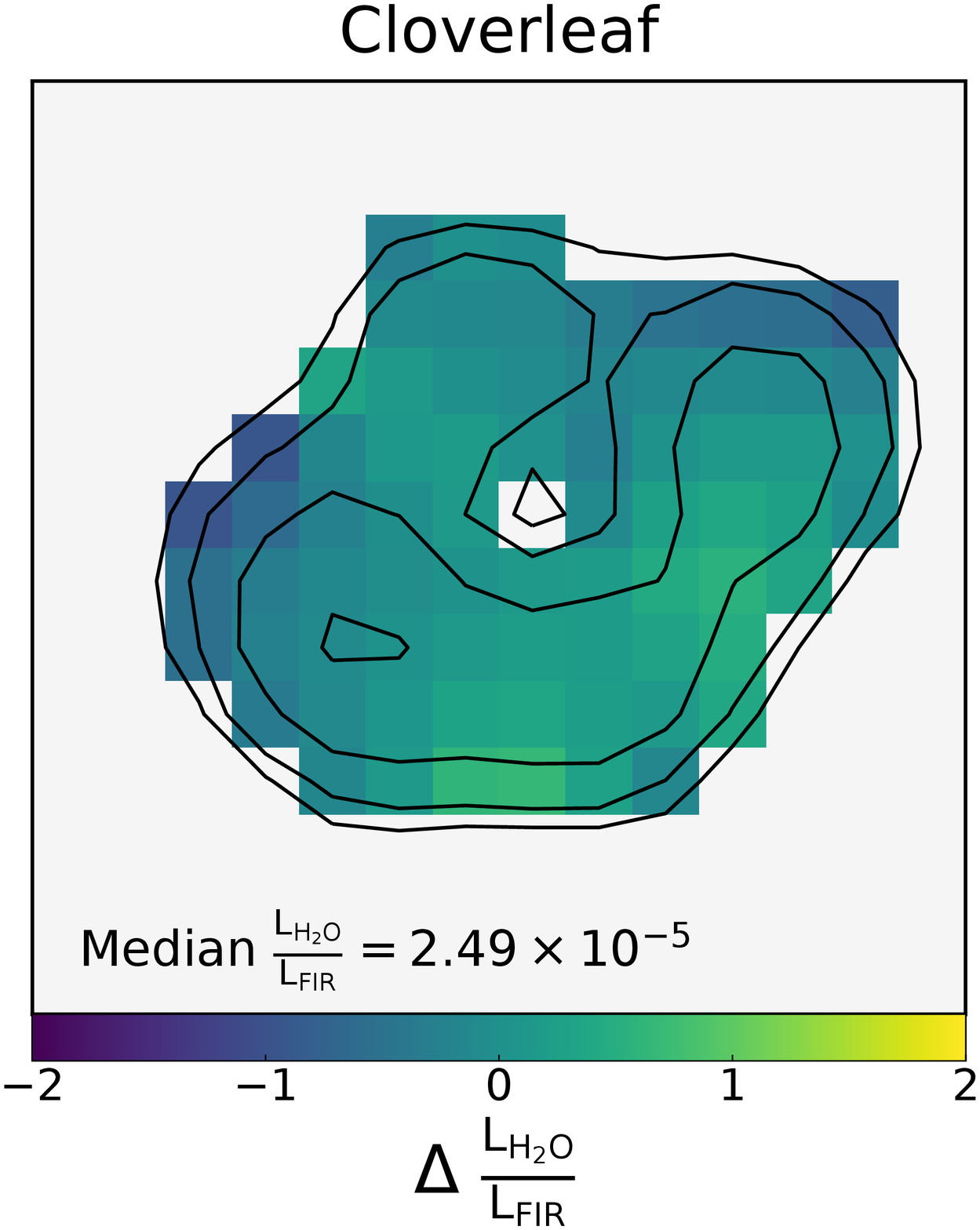}
	\end{tabular}
    \caption{The normalized deviation of \ratio\ in each pixel from the median value i.e. $\Delta\  \frac{\rm L_{\rm H_{2}O}}{\rm L_{\rm FIR}}$ is shown for each source. The contours in black correspond to continuum emission at [3,5,10,20,40,80 ...] x $\sigma$ where $\sigma$ is the RMS noise in continuum map. Values around zero are closer to the median value in that source. The deviation in all the sources is within 10\%.}
    \label{ratioimages}
\end{figure*}

To understand in which regions of the galaxy \water\ best traces \lfir\ in the image plane at resolved pixel scales, we plot the relative deviation of \ratio\ in Figure \ref{ratioimages}. Here, $\Delta\  \frac{\rm L_{\rm H_{2}O}}{\rm L_{\rm FIR}}$ corresponds to [ratio$^{i}$ - ratio$_{med}$] / ratio$_{med}$ where ratio$^{i}$ is \ratio\ of i$^{th}$ pixel and ratio$_{med}$ is the median value of \ratio\ in each source. The closer the $\Delta\  \frac{\rm L_{\rm H_{2}O}}{\rm L_{\rm FIR}}$ value to 0, the closer it is to the median value which implies that \water\ is well correlated with FIR in those pixels. As seen in the images, the correlation is stronger in regions with good signal-to-noise and the ones which deviate the most from $\Delta\  \frac{\rm L_{\rm H_{2}O}}{\rm L_{\rm FIR}}$ are at the edges with low signal-to-noise. Moreover, the bright regions are multiple images of the same region in the source (due to gravitational lensing) and as expected, we see that \ratio\ is the same in these regions. This suggests that water emission faithfully traces the FIR luminosity on resolved scales. 

The almost linear correlation between \lwater\ and \lfir\ at the resolved scales in the galaxies (Figure \ref{mainplot}B and Equation \ref{eqn:lh2o_lfir}) allows us to calibrate the star formation rate as a function of \lwater\ for high-redshift intense star forming regions. The caveat, as discussed previously in Section \ref{sec:intro}, is that the SFR calibration from \lfir\ can not be applied to certain environments such as regions around AGN and resolved SFR depends on variations in dust temperature and opacity. However, we assume a uniform temperature and opacity distribution across the sources as we do not have more resolved continuum observations around the peak of the SED. The SFR is generally estimated from \lir\ scaling relations discussed in \citet{kennicutt12}:
\begin{equation}
\rm SFR\ [M_{\odot}/yr] = 1.47 \times 10^{-10}\ L_{IR}\ [L_{\odot}]
\end{equation}
To convert \lir\ to \lfir, we use $\langle$\lir/\lfir $\rangle$ $\sim$1.38 obtained from the SEDs of SPT0529-54, SPT0532-50, SPT0538-50 and the Cloverleaf. This value is similar to $\langle$\lir/\lfir $\rangle$ in the low-redshift galaxies which is $\sim$1.29 \citep{soifer87}. Using the \lir\ to \lfir\ conversion from the SPT sources and the relation between \lwater\ and \lfir\ given by Equation~\ref{eqn:lh2o_lfir}, we calibrate SFR using \lwater\ at resolved galaxy scales as:
\begin{equation}
\rm SFR\ [M_{\odot}/yr] = 7.35^{+5.74}_{-3.22} \times 10^{-6}\ L_{H_{2}O}\ [L_{\odot}]
\end{equation}
We have shown that \lwater\ is well correlated with the FIR continuum at resolved scales. As mentioned previously, this calibration is applicable to high-redshift intense star forming regions assuming no spatial variations in temperature and optical depth. A similar analysis using resolved continuum observations at the peak of the SED has to be performed to obtain a more accurate SFR calibration from \lwater, which could then be used instead of the observationally expensive \lir.	
We note that resolved FIR continuum observations at the peak of the SED (even at these redshifts) are observationally expensive, while \water\ is a bright line and easily observable in high-redshift sources with ALMA. In addition to providing an alternative to the expensive FIR continuum observations, \water\ additionally provides kinematics of the star forming regions.

\subsection{Effect of AGN}
Aside from the SPT sources, which are dominated by star formation and show no evidence of an AGN, our sample also includes the Cloverleaf quasar, a well characterized AGN at $z$ = 2.56.

The higher transitions of \water\ (E$_{up} \ge$ 400 K) are mainly excited by absorption of short wavelength far-infrared photons ($\le$ 50 $\mu$m) emitted by the hot dust surrounding the AGN. 
Modeling of a lensed quasar, APM 08279+5255 at z $\sim$3.9 \citep{vanderwerf11}, showed that the higher \water\ transitions are arising from the compact central region with \tdust\ $\sim$200 K. 
The AGN contributes less to the J $\le$ 3 excitations (mainly excited by 75  $\mu$m and 101 $\mu$m photons) in the warm regions but does contribute significantly to the total \lir. 
This results in APM 08279+5255 lying low on the $\rm L_{H_{2}O}$/$\rm L_{IR}$ correlation (Figure \ref{mainplot}A, Table \ref{ulirgs_table}). Similar results are found in Mrk 231 where \transe\ is detected \citep{vanderwerf10} and \lwater\ in \transa\ and \transg\ is lower than other ULIRGs without AGN \citep{yang16}.

Previous work \citep{yang13} has shown that although the presence of strong AGN lowers the global $\rm L_{H_{2}O}$/$\rm L_{IR}$ ratio, there does not appear to be a significant effect of AGN on \water\ emission lines. This can be seen from Figure \ref{mainplot}A. Our spatially-resolved analysis of the Cloverleaf quasar in Figure \ref{mainplot}B indicates that \ratio\ remains constant at resolved scales and is similar to ULIRGs even in the presence of an AGN.  
This suggests that the presence of an AGN has little impact on \transa\ excitation -- not just at the global scale but also down to kiloparsec scales.

\subsection{Correlation of $L_{H_{2}O}$/$L_{FIR}$ with physical properties} \label{sec:physicalprop}
The strength of \water\ emission depends on a number of physical properties such as \tdust, \water\ column density and continuum opacity ($\tau$) which are better constrained through modeling of multiple \water\ excitations. As we observed only a single transition, we now investigate to what extent the global variations in \ratio\ we observe correlate with other properties we constrain such as $\rm \lambda_{max}$ (the wavelength in rest frame at which the dust SED peaks) and gas mass density ($\rm \Sigma_{gas}$). We estimate these by using the available photometry and values from the literature (Table \ref{paramtable}). 

\begin{figure*}[h!t]
\hspace*{-1.8cm}
	\begin{tabular}{cccc}
	\includegraphics[trim={0cm 0 0.0cm 0},clip,width=1.00\textwidth]{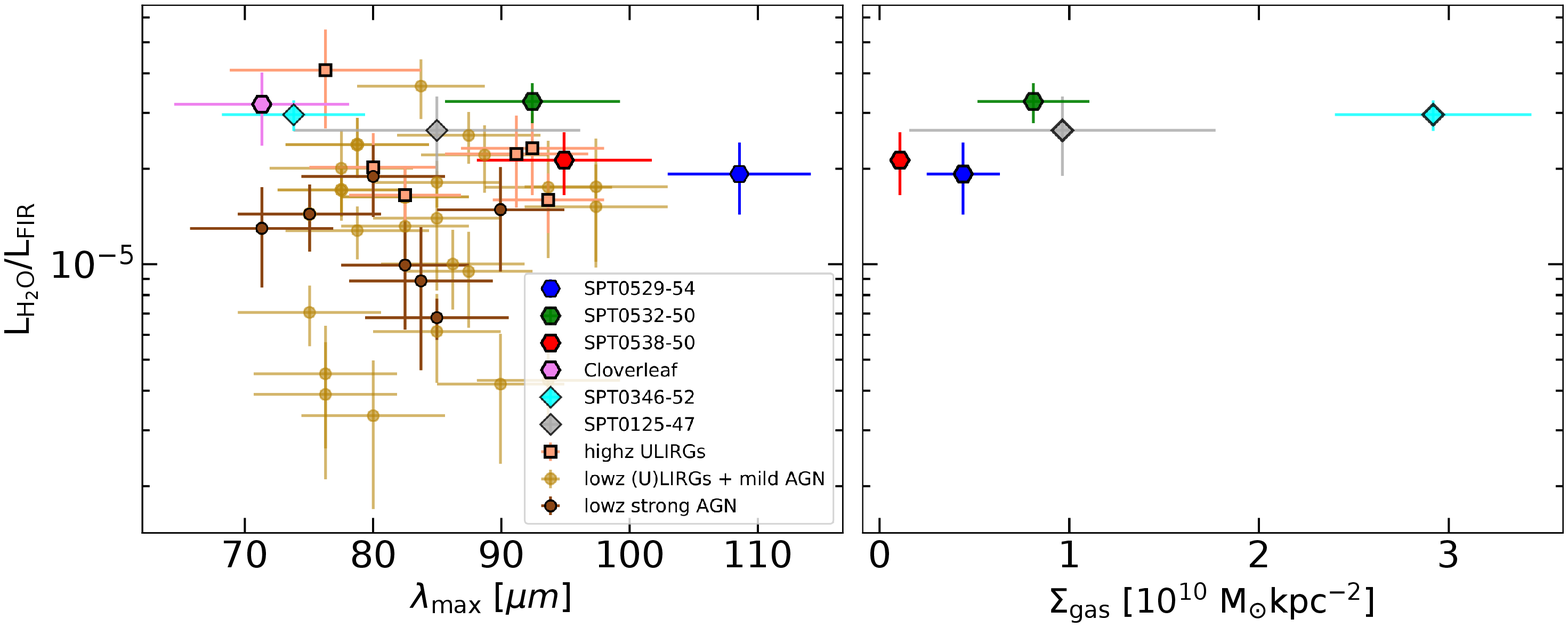}
	\hspace{0.0 cm}
	\end{tabular}
    \caption{\textbf{Left (A):} Correlation between global \ratio\ and $\rm \lambda_{max}$, the rest frame wavelength at which dust SED peaks. \textbf{Right (B):} Global \ratio\ is plotted as a function of gas surface density, $\Sigma_{\rm gas}$ in units of M$_{\odot} \rm kpc^{-2}$. The gas masses for SPT0529-54 and SPT0532-50 are taken from \citet{bothwell17}. SPT0538-50 and other two SPT sources (SPT0125-47 and SPT0346-52) are detailed in \citet{aravena16}. The intrinsic SPT source sizes obtained from lens modeling can be found in \citet{spilker16}. As shown in the plots, the variation in \ratio\ is uncorrelated with either the effective temperature of the dust SED or the gas surface density.}
    \label{paramfig}
\end{figure*}

The correlation of \ratio\ with $\rm \lambda_{max}$ can be interpreted as a correlation with the dust temperature or \lfir\ surface density. Since, \transa\ is mainly excited by the FIR radiation, we are interested in understanding the correlation with dust temperature. We use $\rm \lambda_{max}$ as it is a more direct observable than dust temperature, which is degenerate with optical thickness \citep[e.g.][]{papadopoulos10}. In Figure \ref{paramfig}A, there is no correlation between  \ratio\ and $\rm \lambda_{max}$, consistent with previous results on low-redshift galaxies \citep[Table 2 in][]{yang13} where no relationship is observed between $\rm L_{H_{2}O}$/$\rm L_{IR}$ and S$_{60\mu m}$/S$_{100\mu m}$, a dust temperature indicator (these wavelengths are used to fit the SED in our analysis for local sources). 

In Figure \ref{paramfig}B, we plot \ratio\ as a function of $\rm \Sigma_{gas}$ in an effort to understand whether collisions significantly affect the \water\ excitation. $\rm \Sigma_{gas}$ is calculated by dividing gas mass by the effective area of the source. The gas mass values from \citet{bothwell17} (estimated using \ci\ observations) in SPT0529-54 and SPT0532-50 and from \citet{aravena16} for SPT0538-50, SPT0125-47 and SPT0346-52 (estimated using CO$(1-0)$ observations) are used. The source properties are detailed in \citet{spilker16}, where lens modeling of 870 $\mu$m dust continuum is performed by assuming a single or multiple Sersic source profiles. Using these values, the area ($A_{\rm eff}$) under a Sersic profile is calculated. This method might overestimate $\rm \Sigma_{gas}$ but since the sizes are within a factor of $\sim$2, the overestimated value might only be by a factor of few.
Moreover, the CO (gas) sizes can be larger than that of the infrared emission \citep{spilker15,tadaki17,calistrorivera18,dong19}. For a simple calculation, we assume that the dust and gas sizes are similar. All the values are given in Table \ref{paramtable}. As seen in Figure \ref{paramfig}B, there is no observed correlation of \ratio\ with $\rm \Sigma_{gas}$.

The dust opacity at 100 $\mu$m, the wavelength at which \transa\ is excited, could also affect the intensity of the line and could explain the spread in the \lwater$-$\lfir\ correlation. The slightly super-linear \ratio\ correlation where the increase in \water\ line emission is faster than \lfir\ \citep[e.g.][]{omont13,yang16} could be because of the increase in $\tau_{100}$ (dust opacity at 100 $\mu$m) with the increase in \lfir\ \citep{gonzalezalfonso14b} in turn  enhancing \lwater\ because of photon trapping. In a high $\tau_{100}$ medium, the 100 $\mu$m photons are trapped and scattered thereby increasing the local radiation field. This \mbox{amplifies} the \transa\ pumping and hence the \transa\ line photons.
We do a simple estimation of $\rm \tau_{100}$ in the three SPT sources using the equation from \citet{yang16} where $\rm \tau_{100}$ is given by:

\begin{equation}
\rm \tau_{100} = \kappa_{100} \frac{M_{dust}}{2 \pi r^{2}}
\end{equation}
$\kappa_{100}$ is the dust absorption opacity at 100 $\mu$m and r is the radius of the source at submm wavelength. We use $\rm \kappa_{\lambda}$ = 2.92 $\times\ 10^{5} (\lambda / \mu m)^{-2}$ $\rm cm^{2}g^{-1}$ \citep{li01} at rest wavelength $\lambda$ and dust mass ($\rm M_{dust}$) given by:
\begin{equation}
\rm M_{dust} = \mu^{-1} \frac{D_{L}^{2}S_{\nu}}{(1+z_{s})\kappa_{\lambda}[B_{\nu}(T)-B_{\nu}(T_{CMB})]} 
\end{equation}
Here, $\rm S_{\nu}$ is the flux density at observed frequency, $\rm z_{s}$ is the redshift of the source, $\rm D_{L}$ is the luminosity distance, $\rm B_{\nu}(T)$ is the Planck function at rest frequency (described in Section \ref{sec:lfir}) and $\mu$ is the magnification of the source. From the lens model parameters derived in \citet{spilker16}, we estimate $\rm \tau_{100}$ $\sim$0.34, 1.36 and 0.46 for SPT0529-54, SPT0532-50 and SPT0538-54 respectively. The higher value of $\rm \tau_{100}$ in SPT0532-50 could be enhancing the \water\ luminosity above the average value. However, several physical factors like \tdust, opacity, \water\ abundance, \textit{et cetera}, can also influence the intensity of the line.

To summarize, the global variations in \ratio\ are not observed to be affected by the physical properties of the galaxy such as $\rm \lambda_{max}$ and  $\rm \Sigma_{gas}$. Large dust opacity at 100 $\mu$m might enhance \lwater\ due to photon trapping. Modeling with multiple transitions will give a better understanding of the factors influencing the correlation between \lfir\ and \lwater.

\begin{table*}[htp]
\centering{}
\caption{Observed physical properties in high-redshift ULIRGs}
\label{paramtable}
\begin{tabular}{c c c c c c}
\hline\hline
Source & $\rm \lambda_{max}$ & $\rm M_{gas}$ & $\rm A_{eff}$ & $\rm \Sigma_{gas}$ & Reference\\
 {} & [$\mu$m] & [$\rm 10^{10} L_{\odot}$] & [$\rm kpc^{2}$] &  [$\rm 10^{10} M_{\odot} \rm kpc^{-2}$]\\ 
\hline
 SPT0529-54  & 108.57 $\pm$ 5.59 & 4.05 $\pm$ 0.90 & 9.17 $\pm$ 3.46 & 0.44 $\pm$ 0.19  & \cite{bothwell17}\\
 SPT0532-50 & 92.43 $\pm$ 6.83 & 6.05 $\pm$ 1.71 & 7.46 $\pm$ 1.69  & 0.81 $\pm$ 0.29 &\cite{bothwell17}\\
 SPT0538-50 & 94.91 $\pm$ 6.83 & 1.7  $\pm$ 0.3 &  15.71 $\pm$ 3.05 & 0.11 $\pm$ 0.03 &\cite{aravena16} \\
 SPT0125-47  & 84.98 $\pm$ 11.17 & 11.5 $\pm$ 1.0 & 11.93 $\pm$ 9.93  & 0.96 $\pm$ 0.81 &\cite{aravena16} \\
 SPT0346-52 & 73.80 $\pm$ 5.59 & 8.2 $\pm$ 0.6 & 2.81 $\pm$ 0.45 & 2.92 $\pm$ 0.52 &\cite{aravena16} \\
 Cloverleaf  & 71.32 $\pm$ 6.83 & - & - & - & - \\
 
\hline\hline
\multicolumn{6}{p{\textwidth}}{NOTE. -  $\rm \lambda_{max}$, the rest frame wavelength at which dust SED peaks, is estimated from the modified blackbody fit to photometry using MCMC algorithm. Gas mass ($\rm M_{gas}$) is taken from the references shown in the last column. The Sersic area ($\rm A_{eff}$) is calculated from the best fit source parameters from lens modeling \citep{spilker16}. $\rm \Sigma_{gas}$ is the gas surface density.}
\end{tabular}
\end{table*}

\subsection{$H_{2}O$ and CO} \label{subsec:co}
CO$(6-5)$ traces relatively dense gas (with critical density of H$_{2} \sim10^{5}~\rm cm^{-3}$) in molecular clouds, although not as dense as HCN or HCO$^{+}$ \citep{shirley15,bethermin18}. The high$-$J CO lines are therefore found to be correlated with the far infrared field in these star forming regions \citep[Figure 1 in][]{liu15}. Here, we investigate this correlation in the context of the \lwater$-$\lfir\ relation. We make use of the spatially and spectrally resolved observations of mid$-$J CO$(6-5)$ in SPT0529-54, SPT0532-50 \citep{dong19} and supplement these data with observations of the two other SPT sources SPT0346-52 \citep{apostolovski19} and SPT1247-50 (which is not detected in \transa) from \citet{dong19}. The imaging of the CO data is similar to that described in Section \ref{dataanalysis}. 
The mask used to select the pixels recovers $93$$-$$100\%$ of the CO emission, depending on the source.

\begin{figure*}[h!t]
\hspace*{-1.9cm}
	\begin{tabular}{ccc}
	\includegraphics[trim={0cm 0 0.0cm 0},clip,width=1.08\textwidth]{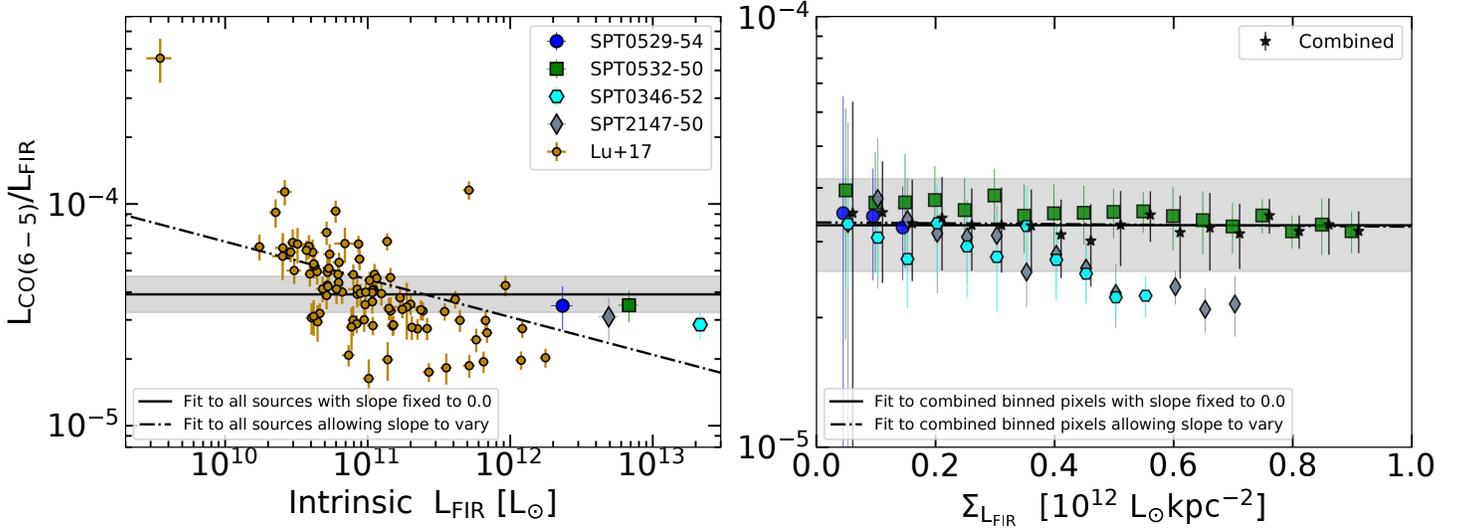}
	\hspace{0.0 cm}
	\end{tabular}
\caption{\textbf{Left (A):} Global $\rm L_{CO(6-5)}$/$\rm L_{FIR}$ as a function of \lfir. The local LIRGs are shown as yellow data points taken from \citet{lu17}. The high redshift ULIRGs are represented by the SPT sources. The \textit{thick black} line is a fit to all the sources by fixing the slope to zero with the 1$\sigma$ error shown as the grey region. The \textit{dot-dashed} line is the fit by allowing the slope to vary. From the plot, we see that the correlation is almost linear. \textbf{Right (B):} Resolved $\rm L_{CO(6-5)}$/$\rm L_{FIR}$ as a function of surface brightness in units of $\rm L_{\odot} kpc^{-2}$. Each data point is the value of pixels binned within 0.05 x $\rm 10^{12}$ $\rm L_{\odot}kpc^{-2}$. The fits are to the combined binned pixels shown in black.
The correlation within the sources follow similar pattern as the global values in the left plot and is also nearly linear. This plot along with Figure \ref{mainplot} suggests that \water\ is as good a tracer of the far infrared radiation as CO$(6-5)$.}
    \label{colfirfig}
\end{figure*}

Figure \ref{colfirfig} shows $\rm L_{CO(6-5)}$/$\rm L_{FIR}$ as a function of \lfir\ similar to Figure \ref{mainplot}. Figure \ref{colfirfig}A contains the global integrated values in local luminous infrared galaxies from \citet{lu17} and high-redshift SPT sources. Figure \ref{colfirfig}B shows the resolved correlation between CO$(6-5)$ and \lfir\ (the binning procedure is similar to that described in Section \ref{subsec:lh2olfir}) where the resolved \lfir\ is estimated using the continuum around the CO$(6-5)$ line. CO$(6-5)$ is observed to have an almost linear correlation with \lfir\ both at global and resolved scales, similar to \water. 

The spectra of CO$(6-5)$ and \transa\ in SPT0532-50 and SPT0346-52 (with a good detection of both the lines) shows that CO has a FWHM consistent with \water\ within the errors (Figure \ref{coh2ofig}A). This may indicate that both the lines are emitted from similar regions in the galaxy \citep[See also][]{omont13,yang16,liu17}. It can further be seen from the spatial distribution comparison in the image plane (Figure \ref{coh2ofig}B). 
This agrees with the results in \mbox{\citet{Yang19}}, where they find similar spatial distribution and also similar kinematic structure between CO$(6-5)$ and \transg\ in G09v1.97.

While \transa\ excitation is due to FIR pumping mechanism and depends mainly on the radiation field density, the CO excitation is due to collisions with the H$_{2}$ molecules. Hence, CO intensity increases with increase in the gas density and temperature \citep[e.g.][]{narayanan14}. The mid and high$-$J CO lines (J = $6-5$ and above) are shown to have increasingly sub-linear slopes with \lfir\ which suggests that denser and much warmer gas than star forming regions (possibly due to shocks or turbulence) is needed for CO excitation \citep{narayanan08,greve14}. Thus, variations in star formation efficiency, or other physical properties within the molecular gas such as the shape of the density probability distribution function (PDF) and the median density of the gas within and between galaxies might \mbox{affect} $\rm L_{CO(6-5)}$/$\rm L_{FIR}$ more strongly than \ratio. 
Although the mid-J $\rm L_{CO}$/$\rm L_{IR}$ ratio is not expected to be enhanced in galaxies with supernovae or stellar wind driven shocks, NGC 6240 shows a higher ratio \citep{lu17}.
This suggests that \water\ is an intrinsically better tracer of the far infrared field than CO$(6-5)$. 
To confirm this result, we need a larger sample of sources across a broad range in \lfir\ to compare \water\ and CO (and other dense gas traces such as HCN) to determine which one is an empirically better tracer of star formation.

\section{Summary and Conclusion} \label{sec:conclusion}
We observed \transa\ 987.927 GHz line in SPT0529-54 ($z$ = 3.369), SPT0532-50 ($z$ = 3.399) and SPT0538-50 ($z$ = 2.782) with ALMA. We also include the Cloverleaf quasar at $z$ = 2.558 to compare with the star forming galaxies.
The observational results and conclusions from this analysis are:
\begin{itemize}
\itemsep0.2em 
\item{\lwater\ is empirically correlated with \lfir\ over more than three orders-of-magnitude from low-redshift LIRGs to high-redshift ULIRGs}
\item{The relationship between \lwater\ and \lfir\ stays \mbox{linear} even at resolved scales within individual galaxies with average \ratio\ = 2.76$^{+2.15}_{-1.21} \times$ 10$^{-5}$}
\item{This linear correlation holds even in the presence of a strong AGN in the Cloverleaf quasar}
\item{We present \transa\ as a resolved SFR calibrator for high-redshift intense star forming regions assuming a single temperature and opacity across the source
\begin{equation*}
\rm SFR\ [M_{\odot}/yr] = 7.35^{+5.74}_{-3.22} \times 10^{-6}\ L_{H_{2}O}\ [L_{\odot}]
\end{equation*}}
\item{There is no observed correlation of \ratio\ with $\rm \lambda_{max}$, the wavelength at which SED peaks or $\rm \Sigma_{gas}$, the gas mass surface density. The dust opacity at 100 $\mu$m ($\tau_{\rm100}$), on the other hand, may influence \lwater\ due to photon trapping. However, the current sample is too small to give any definite result}
\item{\transa\ is intrinsically a better tracer of \lfir\ than CO$(6-5)$}. A larger sample size is needed to confirm this result
\end{itemize}
This work shows that \transa\ traces \lfir\ at resolved $\sim$kiloparsec scales in high-redshift galaxies with intense star forming regions while assuming a single temperature and dust opacity across the source. In order to validate these assumptions and obtain a more accurate SFR calibration, we need resolved continuum observations around the peak of the SED. We also need to perform similar analysis on less luminous galaxies (\lfir\ $<$ 10$^{12}$ L$\odot$) to extend the SFR calibration. Future work will involve detailed lens modeling of the sources with a pixellated lens model \citep{hezaveh16}. In the future, it would also be interesting to compare and model multiple resolved \water\ lines with other dense gas tracers.

\section{Acknowledgments}
This paper makes use of the following ALMA data: ADS/JAO.ALMA \#2015.1.01578.S, \#2016.1.01554.S, \#2012.1.00844.S, \#2012.1.00175.S and \#2011.0.00957.S. The SPT is supported by the NSF through grant PLR-1248097, with partial support through PHY-1125897, the Kavli Foundation and the Gordon and Betty Moore Foundation grant GBMF 947. S.J., J.D.V., D.P.M. and K.C.L.  acknowledge support from the US NSF under grants AST-1715213 and AST-1716127. S.J. and K.C.L acknowledge support from the US NSF NRAO under grants SOSPA5-001 and SOSPA4-007, respectively.
J.D.V. acknowledges support from an A. P. Sloan Foundation Fellowship. D.N. was supported in part by NSF Award AST-1715206 and HST Theory Award 15043.0001. ALMA is a partnership of ESO (representing  its  member  states), NSF (USA) and NINS (Japan), together with NRC (Canada), MOST and ASIAA (Taiwan), and KASI (Republic of Korea), in cooperation with the Republic of Chile. The Joint ALMA Observatory is operated by ESO, AUI/NRAO and NAOJ. This research has made use of NASA\textsc{\char13}s Astrophysics Data System.

\bibliography{spt_smg_h2o}

\section{Appendix: A1} 

\begin{table*}[htp]
\centering
\caption{Observed properties in high-redshift ULIRGs}
\label{ulirgs_table}
\begin{tabular}{c c c c c c c c}
\hline\hline
Source & z & $\mu$ & $\lambda_{max}$ & \lfir/$\mu$ & \lwater/$\mu$ & \ratio & Reference\\
{} & {} & {} & [$\mu$m] & [10$^{12}$ L$_{\odot}$] & [10$^{8}$ L$_{\odot}$] & [10$^{-5}$] & {} \\ 
\hline
SPT0125-47 & 2.5148 & 5.467 $\pm$ 0.120 & 84.98 $\pm$ 11.17 & 19.40 $\pm$ 4.37 & 5.12 $\pm$ 0.87 & 2.64 $\pm$ 0.74 & Appendix A2 [\ref{sec:appendix2}] \\
SPT0346-52 & 5.6559 & 5.570 $\pm$ 0.117 & 73.80 $\pm$ 5.59 & 21.50 $\pm$ 2.31 & 6.36 $\pm$ 0.24 & 2.96 $\pm$ 0.32 & \citet{apostolovski19}\\
G12.v2.30 & 3.259 & 9.5 $\pm$ 0.6 &  82.49 $\pm$ 4.34 & 8.16 $\pm$ 1.02 & 1.35 $\pm$ 0.27 & 1.65 $\pm$ 0.39 & \citet{omont13}\\
NAv1.195 & 2.951 & 4.1 $\pm$ 0.3 &  93.67 $\pm$ 4.34 & 10.25 $\pm$ 1.38 & 1.63 $\pm$ 0.27 & 1.59 $\pm$ 0.34 & \citet{yang16}\\
SDP11 & 1.786 & 10.9 $\pm$ 1.3 &  91.18 $\pm$ 5.59 & 2.60 $\pm$ 0.63 & 0.58 $\pm$ 0.12 & 2.22 $\pm$ 0.71 & \citet{yang16}\\
NBv1.78 & 3.111 & 13.0 $\pm$ 1.5 &  80.01 $\pm$ 4.97 & 4.65 $\pm$ 0.79 & 0.94 $\pm$ 0.21 & 2.02 $\pm$ 0.57 & \citet{omont13}\\
SDP17 & 2.305 & 4.9 $\pm$ 0.7 &  92.43 $\pm$ 5.59 & 7.47 $\pm$ 1.65 & 1.73 $\pm$ 0.32 & 2.32 $\pm$ 0.67 & \citet{omont13}\\
HFLS3 & 6.337 & 2.2 $\pm$ 0.3 & 76.28 $\pm$ 7.45 & 13.47 $\pm$ 3.74 & 5.51 $\pm$ 1.12 & 4.09 $\pm$ 1.41 & \citet{riechers13} \\
APM08279 & 3.9 & 4.0 &  & 50.0 $\pm$ 12.0 & 6.0 $\pm$ 1.2 & 1.2 $\pm$ 0.3 & \citet{vanderwerf11} \\
 +5255&  &  &  &  &  &  &  \\

\hline\hline
\multicolumn{8}{p{\textwidth}}{NOTE. - For the SPT sources, z and  magnification ($\mu$) are given in \citet{spilker16} and $\lambda_{max}$ (the wavelength in rest frame at which the dust SED peaks) and \lfir\ are estimated by fitting a modified blackbody function to the photometry by fixing $\beta$ = 2.0. \water\ observations of SPT0346-52 is discussed in detail in \citet{apostolovski19}. In HFLS3, photometry is from \citet{riechers13} and magnification is from \citet{cooray14}. The magnification in APM08279+5255 is taken from \citet{riechers09}. For all other sources, $\mu$ and photometry is from \citet{bussmann13}. \lfir\ and $\lambda_{max}$ are estimated by fixing $\beta$ = 2.0 and $\lambda_{0}$ = 100 $\mu$m except in APM08279+5255 where \lfir\ is taken from \citep{beelen06,weiss07}. \lwater\ is taken from the references given in the last column.}
\end{tabular}
 \end{table*}

\section{Appendix: A2} \label{sec:appendix2}
\begin{figure*}[h!t]
\centering
	\begin{tabular}{ccc}
	\includegraphics[trim={0cm 0 0.0cm 0},clip,width=0.40\textwidth]{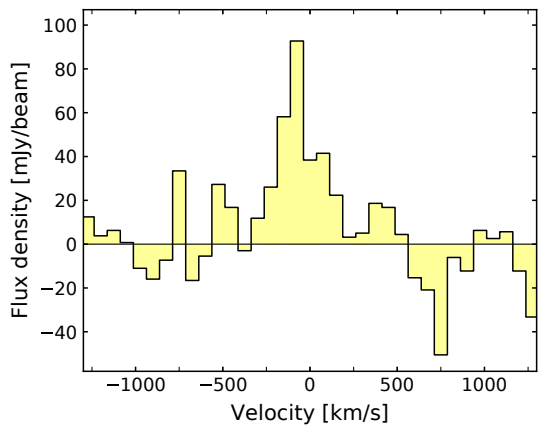}
	\hspace{0.0 cm}
	\end{tabular}
\caption{The spatially integrated spectrum of \transa\ transition with 75 kms$^{-1}$ spectral resolution in SPT0125-47.}
    \label{spt0125_spectrum}
\end{figure*}

\begin{figure*}[h!t]
\hspace*{-1.0cm}
	\begin{tabular}{cccc}
	\includegraphics[trim={0.37cm 10.0 0cm 10.0},clip,width=0.46\textwidth]{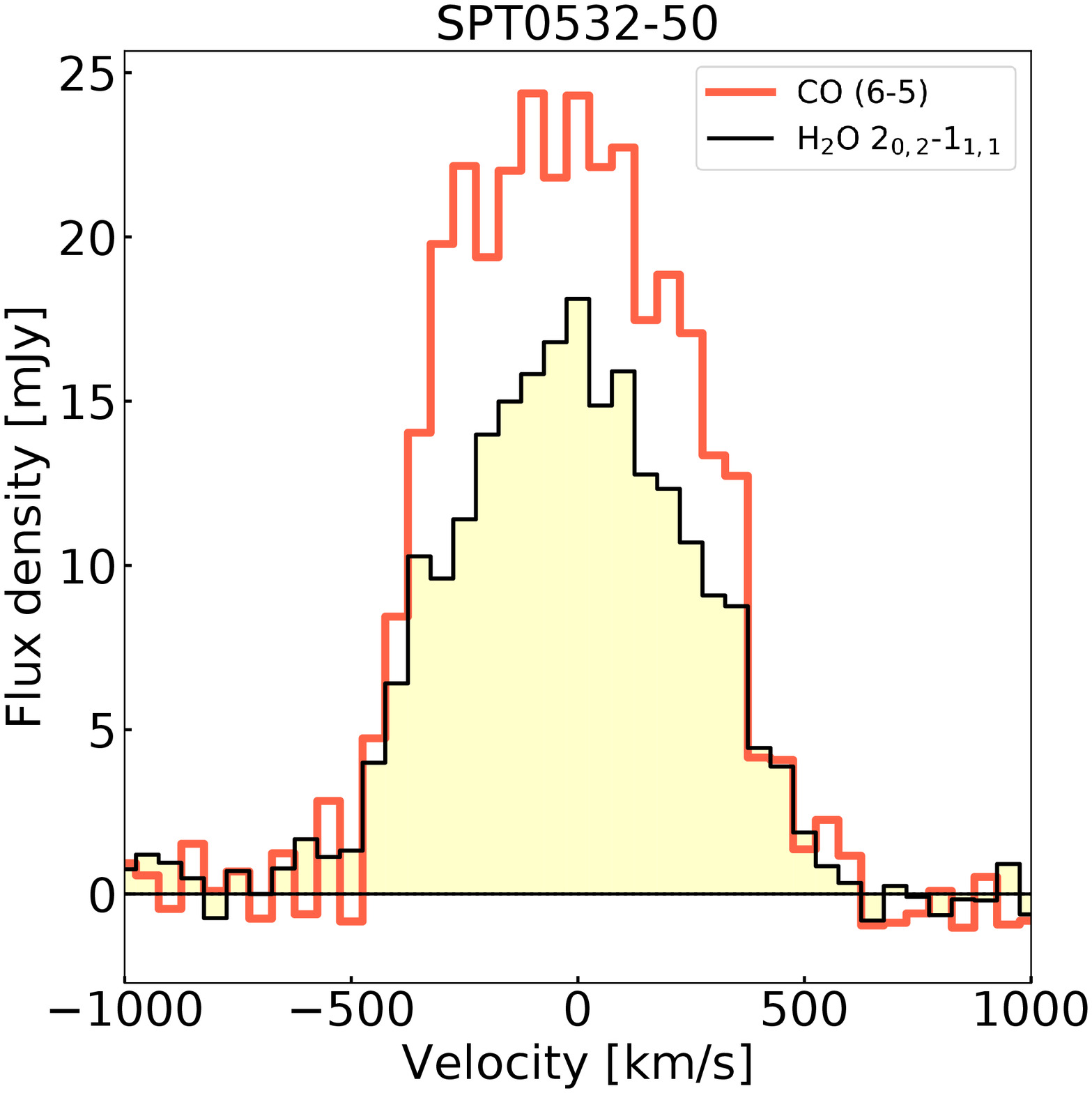}
	\hspace{0.0 cm}
	\includegraphics[trim={0.0cm 5.0cm 0.0cm 5.0cm},clip,width=0.57\textwidth]{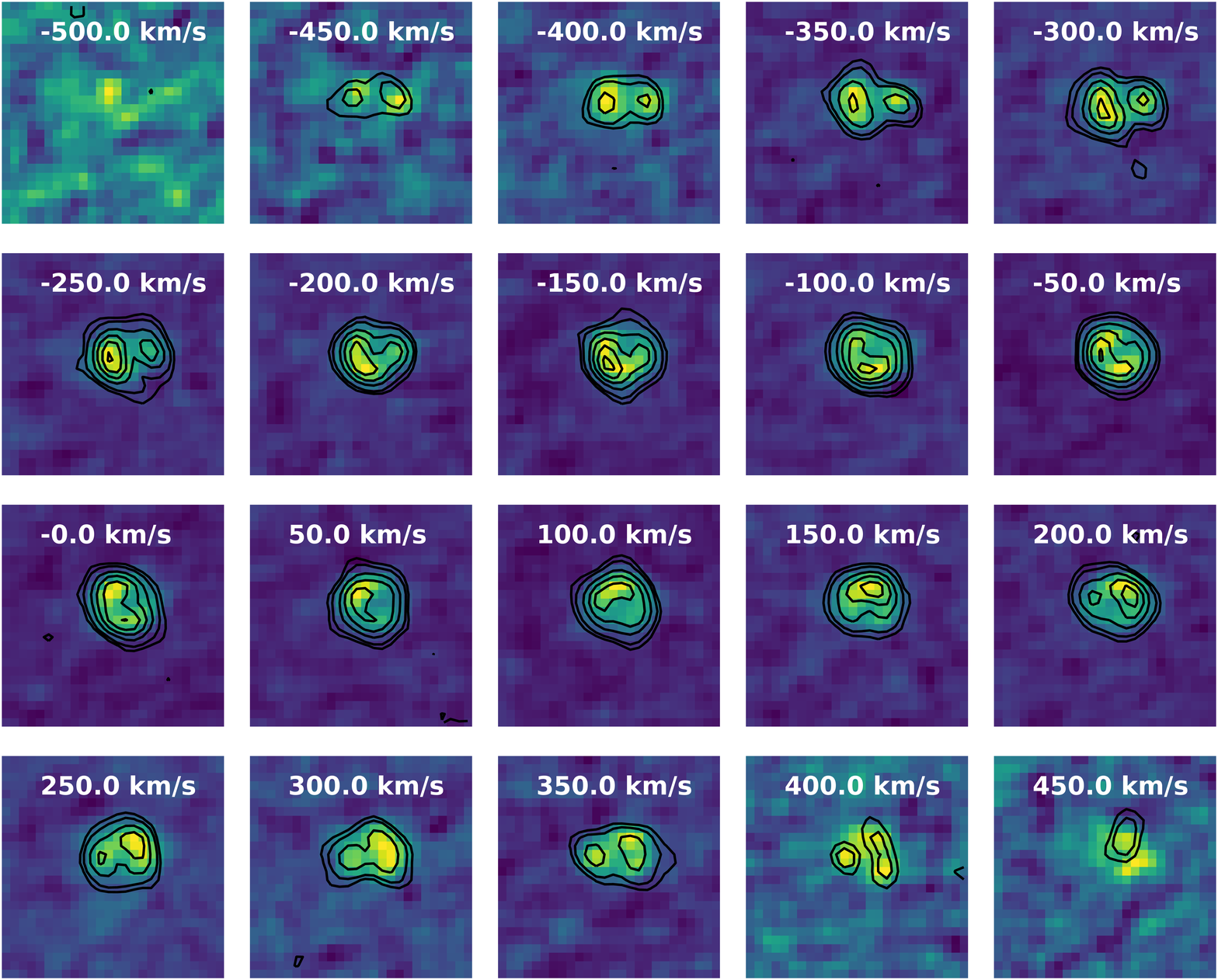}\\\\
	\hspace{-0.5 cm}
	\includegraphics[trim={0.38cm 10.0 0cm 12.0},clip,width=0.48\textwidth]{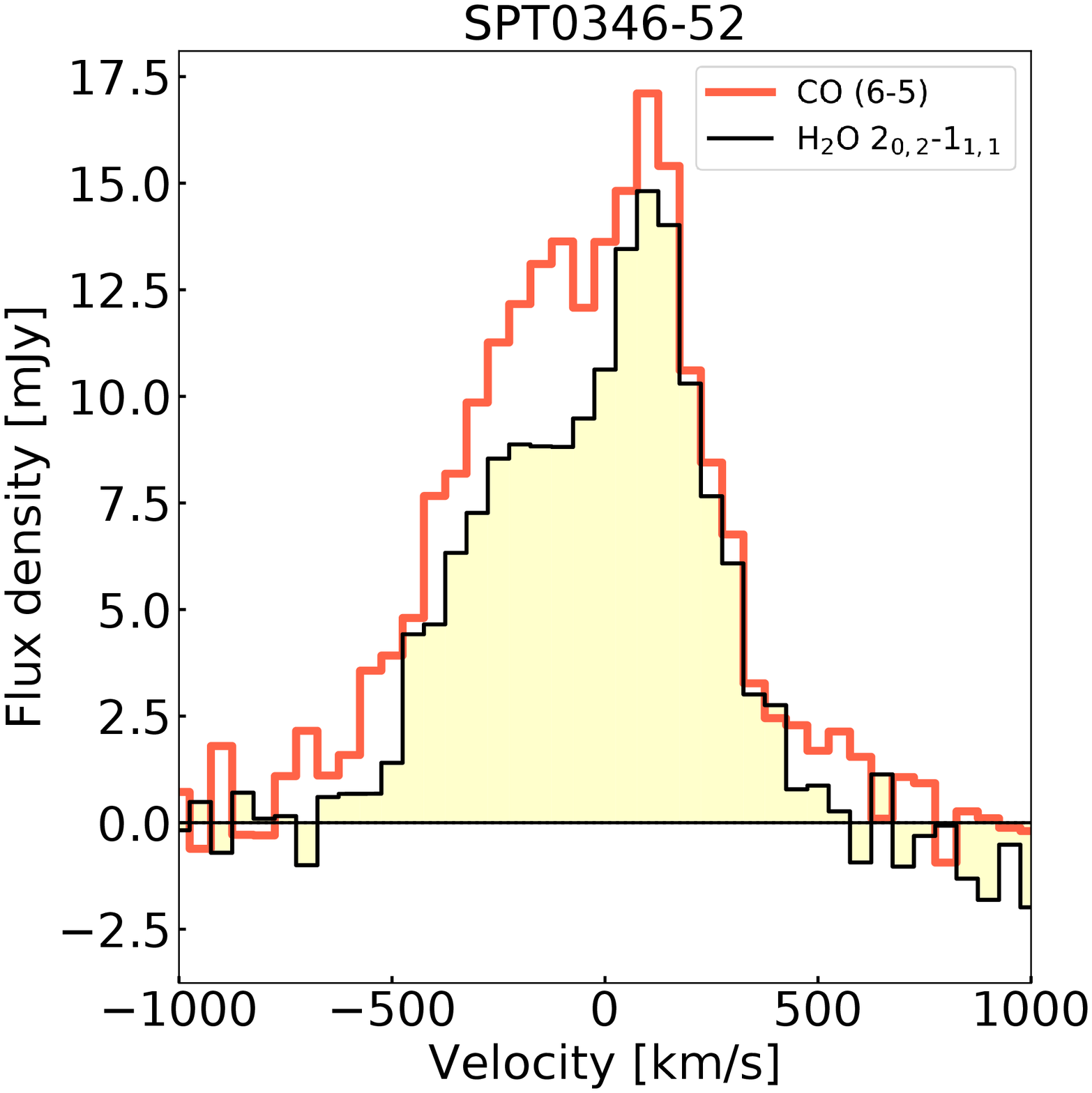}
	\hspace{0.0 cm}
	\includegraphics[trim={0.0cm 5.0cm 0.0cm 5.0cm},clip,width=0.57\textwidth]{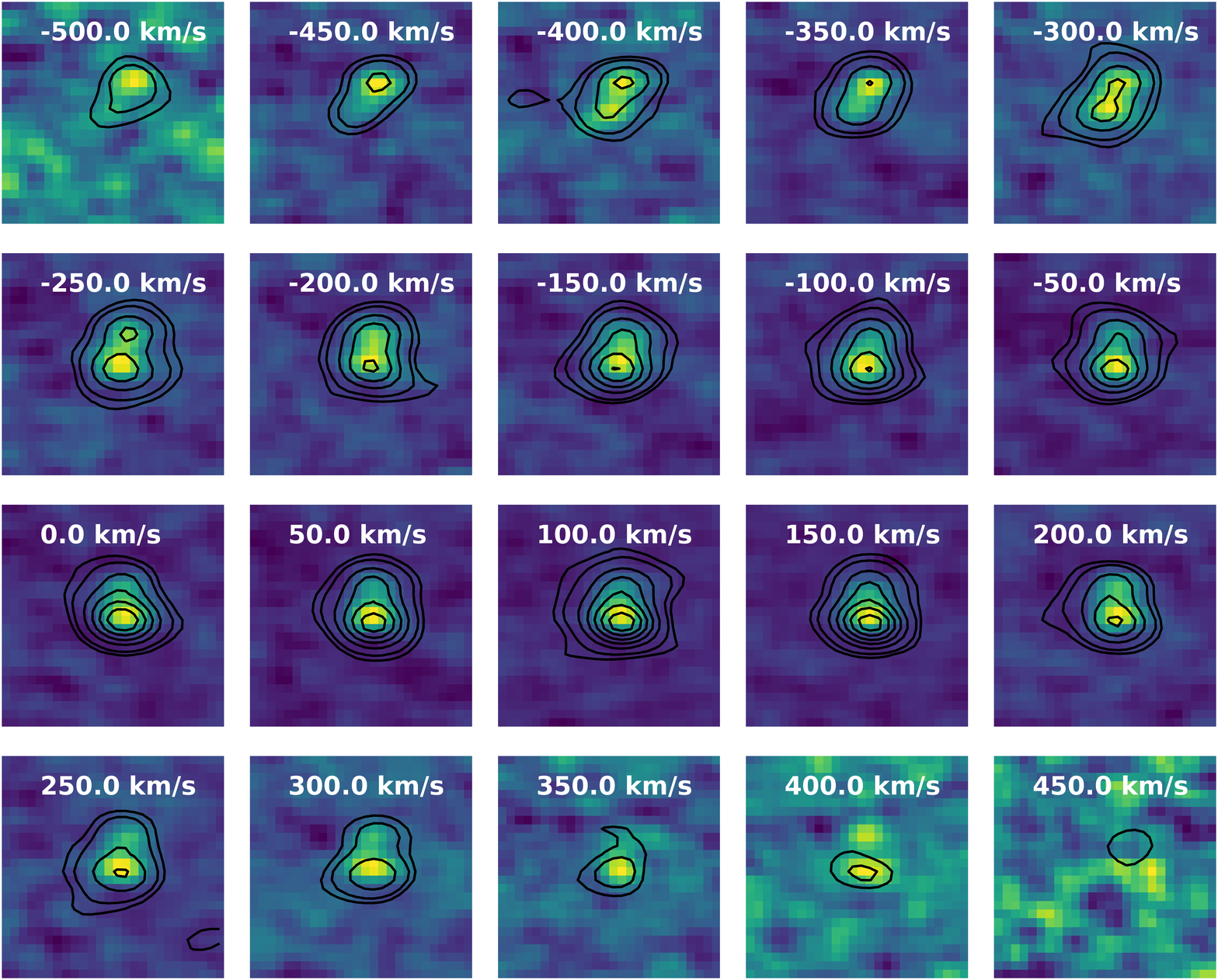}
	\end{tabular}
    \caption{\textbf{Left (A):} Spectra of CO$(6-5)$ and \transa\ in SPT0532-50 and SPT0346-52 integrated over 50 kms$^{-1}$ channels. Both CO and \water\ have similar FWHM. \textbf{Right (B):} The channel map of \transa\ in the background and CO$(6-5)$ contours overlaid in black in both the sources. The contours are at [3,5,10,15 ...] x $\sigma$ where $\sigma$ is the RMS noise in the CO cube. CO$(6-5)$ and \water\ may be tracing similar warm dense regions of the galaxy.}
    \label{coh2ofig}
\end{figure*}

 \begin{center} 
 APEX observations of \water\ in SPT0125-47
 \end{center}
We observed \transa\ ($\nu_{\rm rest}$=987.927 GHz) line in SPT0125-47 at $z$=2.5148 using the APEX-2 receiver of the Swedish Heterodyne Facility Instrument \citep[SHFI;][]{vassilev08} on the Atacama Pathfinder Experiment (APEX). The observations in the shared ESO+Swedish project 092.A-0467 (PI M. Aravena) were done between July and November 2013 in excellent conditions with Precipitatable Water Vapor 0.25$<$PWV$<$0.5\,mm, and a total on-source integration time of 3 hours. We reduced the data using the standard procedures in the IRAM CLASS software. The line is clearly detected (Figure \ref{spt0125_spectrum}) with a line flux of 21.8$\pm$3.7\,Jykm\,s$^{-1}$ and a line width of $\sim$117\,km\,s$^{-1}$. Note that the source is unresolved in the 280\,GHz APEX beam of 22\farcs3.

\section{Appendix: A3} 
\label{sec:appendix3}
\begin{center} 
 Spatial distribution of CO$(6-5)$ and \water\ in SPT0532-50 and SPT0346-52
 \end{center}
From the spatial distribution comparison of CO$(6-5)$ and \transa, we can see that both the lines are tracing similar regions in the velocity space. Although, the source is gravitationally lensed, the similar distribution in the image plane might indicate that they are tracing the same regions of the galaxy in the source plane.

\end{document}